# Exploring Microtask Crowdsourcing as a Means of Fault Localization

Christian M. Adriano, *Member, IEEE,* André van der Hoek, *Member, IEEE*

**Abstract**— Microtask crowdsourcing is the practice of breaking down an overarching task to be performed into numerous, small, and quick microtasks that are distributed to an unknown, large set of workers. Microtask crowdsourcing has shown potential in other disciplines, but with only a handful of approaches explored to date in software engineering, its potential in our field remains unclear. In this paper, we explore how microtask crowdsourcing might serve as a means of fault localization. We particularly take a first step in assessing whether a crowd of workers can correctly locate known faults in a few lines of code (code fragments) taken from different open source projects. Through Mechanical Turk, we collected the answers of hundreds of workers to a pre-determined set of template questions applied to the code fragments, with a replication factor of twenty answers per question. Our findings show that a crowd can correctly distinguish questions that cover lines of code that contain a fault from those that do not. We also show that various filters can be applied to identify the most effective subcrowds. Our findings also presented serious limitations in terms of the proportion of lines of code selected for inspection and the cost to collect answers. We describe the design of our experiment, discuss the results, and provide an extensive analysis of different filters and their effects in terms of speed, cost, and effectiveness. We conclude with a discussion of limitations and possible future experiments toward more full-fledged fault localization on a large scale involving more complex faults.

——————————— ◆ ———————————

## 1 Introduction

Crowdsourcing – the act of taking a task traditionally performed by a designated agent (such as an employee or a contractor) and outsourcing it by making an open call to an undefined but large group of people [1] - has emerged as an alternative approach to accomplish work. While crowdsourcing has been famous for examples such as FoldIt [2], the DARPA red balloon challenge [3], and Wikipedia [4], nowadays platforms such as Amazon Mechanical Turk (MTurk) [5], ideascale [6], and Upwork [7] offer anyone the ability to engage a crowd of workers in a task at hand.

The software industry has taken note of the potential of crowdsourcing [8] and a variety of platforms have emerged through which one can engage different crowds in different kinds of software development work. For instance, TopCoder [9] has hosted over 427,000 software design and development competitions and uTest [10] enlists over 100,000 freelancers to test new apps for device compatibility, actual functionality, and usability. Another example is StackOverflow [11], which is driven by volunteers who have together posted more than 16,000,000 answers to programming questions. Additionally, companies such as Netflix, Microsoft, and Facebook regularly post bug bounties in which anyone can choose to participate [12], [13].

Significant variability exists; for instance, workers can bid for, compete on, or be assigned to a task. Tasks may take days, hours, or minutes. Work can be distributed as a whole, broken down into smaller subtasks, or split into fine-grained microtasks. Workers can be experts in, have knowledge of, or even be unfamiliar with the domain. Because of such variability, several canonical models of crowdsourcing have emerged. The most popular models are peer production (e.g., StackOverflow, Open source software development), competition (e.g., TopCoder, CodeHunt [14]), and microtasking (e.g., CrowdCode [15], uTest). Other models exist such as gamified approaches (e.g., CrowdMine [16], Pipe Jam [17]).

We explore in this paper the microtasking model of crowdsourcing, in which a large task is broken down into numerous, small, and quick to complete microtasks that are distributed to an unknown large set of workers [18], [19]. To date, it can be argued that microtasking has dealt more with the outward side of software (i.e., the user interface and the functionality it exposes) than the inward side (i.e., the code itself and its underlying design). This is certainly true in practice, with platforms such as uTest, UserTesting [20], and TryMyUI [21]. It is mostly true in research as well, although a small handful of exceptions exist in the form of experimental tools such as CrowdCode and Collabode [22].

Our work concerns fault localization [23], and particularly investigates if microtask crowdsourcing could serve to locate faults in software. Clearly, this represents a significant challenge for which various microtasks will need to be orchestrated into an over-arching workflow. As a first step toward such a work-flow, we contribute an experiment that is designed to determine whether a crowd of workers can correctly locate faults in relatively small code fragments (i.e., a few lines of code).

Guided by several preliminary feasibility studies, the experiment described here is rooted in five key decisions:
- We conducted the experiment through Mechanical

————————————————
- *F.A. Author is with the University of California Irvine, Irvine, CA, 92697-3440 E-mail: adrianoc@uci.edu.*
- *S.B. Author is with the University of California Irvine, Irvine, CA, 92697-3440 E-mail: andre@ics.uci.edu.*

- Turk so that we could reach a broad set of workers and, in the process, assess the impact of different types of workers (e.g., professionals, students) on the eventual results.
- Each HIT (a microtask on Mechanical Turk) consisted of three questions about a Java method containing a known fault. Each Java method and associated fault were taken from Defects4J [24] so that the experiment applies to real code and real faults. We selected a total of eight failing Java methods, each containing a single fault.
- All the questions were automatically generated from four templates, such that the approach is generalizable and can be repeated and translated to practice without requiring manual creation of questions.
- Each worker could complete up to eight HITs, with each HIT corresponding to a different Java method and its fault. In this manner, no single worker can dominate the overall results.
- Each question was answered by 20 different workers so that we could analyze how much replication is necessary for a crowd to actually locate the faults.

Of note in the experimental design is that we only selected faults that were contained within a single Java method. Faults that would require inter-procedural analysis, concurrency analysis or other more program-wide examinations were excluded. Addressing more complex faults will require significant extensions, both in the kinds of questions asked and the interface through which the questions are asked – a subject of our future work.

Concerning our results, all eight faults were successfully located. This means that for each of the eight faults, at least one of the questions that covered the faulty lines of code was correctly identified as the location of the fault by the workers' collective answers. However, we also identified trade-offs among speed (time to locate faults), cost (number of workers and number of answers), and effectiveness (how many lines of code a user has to inspect). For instance, the least expensive subcrowd (fewest number of workers and answers) consisted of non-students who scored 100% in the qualification test. Nonetheless, the questions selected by this subcrowd did not locate faults within the fewest lines of code (level of effectiveness). Instead, the most effective subcrowd was obtained by filtering out the answers to certain question types. This subcrowd located all the lines containing the faulty code (10), but also incorrectly selected another 24 lines. This represents 11% of the lines of code of the selected Java methods, hence, a significant overhead for programmers to inspect.

The remainder of the paper is organized as follows. Section 2 provides relevant background material in crowdsourcing and fault localization. Section 3 presents our experimental design, and Section 4 reviews the demographics of the workers who chose to participate. Section 5 provides a detailed analysis of the results. Section 6 discusses our findings and their implications. Section 7 reviews threats to validity. Section 8 describes the related work, and Section 9 concludes with an outlook at our future work.

## 2 Background

In this section, we first discuss relevant background in crowdsourcing in software engineering, detailing the microtask crowdsourcing approach. We briefly review key aspects of fault localization that are relevant to our experiment.

### 2.1 Crowdsourcing in Software Engineering

Crowdsourcing in software engineering happens in many successful ways. Open source software development benefited from the contributions from hundreds and hundreds of volunteers [25], [26], [27]. Software competitions have brought together hundreds of programmers to quickly accomplish complex software tasks [9], [12], [13]. Question & answer sites [11], [28], [29] enabled developers to crowdsource frequent coding problems and receive dozens of solutions in a few minutes [30]. While these examples show how crowdsourcing changed software development, academia has also been exploring crowdsourcing on different types of software development work. For instance, research has explored crowdsourcing of requirement extraction [31], user interface prototyping [32], issue triaging [33], software testing [10], [20], [21], and software verification [16], [17], [34].

Cutting across these examples of crowdsourcing in software engineering, La Toza and van der Hoek recognized several crowdsourcing models [8]: peer-production, competition, gamification, and microtasking. Each model has a way of engaging the crowd and coordinating the individual efforts. In the peer-production model, for instance, crowds work together without any centralized control [25], [11]; in a sense, everyone chooses what to work on and when. In the competition model, as another example, programmers compete for monetary rewards [9], [35], points [14], or just to bid for a job [7], [36]. In the gamified model, individuals perform actions in a game [16], [17], which are translated to actual software development tasks. Another model is microtasking, in which a larger work, e.g., usability testing, is partitioned among many workers [10], [20], [21], who individually test different parts of a software.

Of these crowdsourcing models, the microtasking and the gamified models have been the least explored in software engineering to date. In contrast to the gamified model, microtasking is already a practice in the software industry. However, utilizing microtasks is still challenging for more complex and interdependent software development tasks [37] (e.g., coding and debugging). These limitations explain why microtasking has been mostly adopted to crowdsource work that requires limited context, e.g., usability testing [15].

### 2.2 Microtasking in Software Engineering

The motivations to utilize microtasks as an approach to software development are that, ideally, work is partitioned into tasks that can be executed in a few minutes (short), have all necessary information (self-contained), can be ac-complished independently by different workers (parallelizable), and can be recombined into a larger result [38]. Hence, to develop software through microtasks, a requester has to determine how to partition work, identify all the necessary microtasks, delegate these microtasks efficiently, and guarantee that the final outcome has acceptable quality.

A typical solution to partition work is to break down work along the borders of individual software artifacts (e.g., user stories, test cases, and functions) [39]. For instance, in CrowdCode [15], one function and corresponding test cases are all partitioned to be implemented through multiple and distinct microtasks. This way, while multiple workers write tests for a function, another worker can be adding pseudo-code to the same function or replacing it by the actual code. Similarly, TryMyUI [21] breaks down each test case in a list of microtasks, each corresponding to single actions on a GUI, for instance, to perform a search by a certain keyword.

Identifying all the necessary microtasks varies based on type of software activity. While in software testing and verification the set of microtasks can be established a priori, in specifications, coding or debugging new microtasks are created based on the outcome of previous microtasks. For instance, CrowdCode [15] and VeriWeb [34] generate new microtasks based on the impact of concluded microtasks on work that was already accomplished. In a similar, but manual way, Collabode [22] relies on an "original programmer" to decide which microtasks are necessary at any given time, for example, refactoring microtasks to fix compilation errors caused by changes in method signatures.

The delegation of microtasks and the quality of outcomes are shown to be interdependent issues. For instance, a user study with Collabode [22] showed that when microtasks were delegated in a way that workers could work independently, broken builds lasted longer. Conversely, quality of outcome can also impact how microtasks are delegated. In a controlled experiment with CrowdCode [15], one worker accomplished more than expected for a certain microtask, in particular, the coding of a new function. The consequence was that new microtasks could not be created, hence impacting the delegation of new microtasks. Other studies also reported [40], [41], [42], the presence of workers who perform a disproportionate number of microtasks, thereby biasing the crowdsourcing outcomes.

## 2.3 Fault Localization

When faced with a software failure, programmers search for the corresponding location of the fault by printing statements, setting debugger breakpoints [43], or simply inspecting the source code for possible bugs [44]. Regardless of the approach, programmers typically must inspect many lines of code to locate a single fault. In order to alleviate such manual work, researchers have investigated different fault localization techniques [45], [46], for instance, program slicing [47], [48], algorithmic debugging [49], delta debugging [50], [51], and statistical fault localization [23], [52], [53], [54].

The experiment we report in this paper is most closely related to statistical fault localization. This technique operates by associating each set of source code lines with a level of suspiciousness of containing a fault [23]. Although different formulas have been proposed to compute the level of suspiciousness [23], [53], [54], [55], [56], [57], all techniques are based on how many times failing and passing unit tests execute certain source code lines. Additionally, the levels of suspiciousness are associated with different source code granularities. For instance, the most common granularities are program statement [23], [58], predicate [54], class [59], method [60], blocks [52], [53], and branch or def-use dependency [61].

## 3 EXPERIMENT DESIGN

The goal of our experiment was to evaluate whether faults could be effectively located in terms of time, cost, and effectiveness. However, as a preliminary evaluation, we constrained the experiment to simple bugs, given that that they are located in a single Java method. Our assumption was that if this simple case showed positive results, we would be able to explore more complex bugs in future experiments.

Concerning the current experiment, we adopted a simple workflow, through which we recruited workers and asked them questions about the relation between source code fragments and software failures in different Java methods. We then aggregated the workers' answers in order to identify the source code lines considered to contain the faults.

The next subsections describe how we designed this simple workflow. We relied on five design choices regarding how to: (1) qualify and compensate workers, (2) ask questions about code fragments, and (3) generate, (4) distribute and (5) replicate these questions through microtasks.

### 3.1 Worker Recruitment

While many alternatives exist to recruit workers [7], [62], [63], [64], [65], [66], [67], we chose MTurk because it has a large and diverse set of workers [68], [69]. In this regard, we were particularly interested in reaching professional programmers, hobbyists, and college students, so we could evaluate the experiment outcomes under different levels of worker programming skill.

Note that we did not announce our study on any academic, professional, or crowdsourcing mailing lists or forums. We relied solely on MTurk interface, thereby mitigating distortions in recruitment speed or workforce quality, which might result from workers sharing recommendations online [70].

Any MTurk worker could sign up for the experiment, but we did use a qualification test to ensure that only workers with basic programming skills could participate. We designed four tests with five questions each about program outputs in the face of possible changes in certain lines of code. Workers needed to answer at least three of the questions correctly to be qualified. Moreover, we used four different tests to mitigate answers becoming known. In addition, each worker could take each different test once, thereby we avoided workers retaking tests until they passed.

Upon completion of the HIT, each worker was compensated one dollar per microtask, roughly equivalent to the California minimal wage, with each microtask estimated to five minutes ($9/h). For the qualification test, workers were also compensated one dollar.

### 3.2 Fault localization microtask

Central to the experiment were the microtasks presented to the workers. Each microtask consisted of one question about a possible relation between one code fragment and a software failure. In this way, questions about code fragments enabled us to partition the larger work of fault localization into microtasks.

#### 3.2.1 Use of Questions for Fault Localization

Besides work partitioning, the use of questions also allowed each worker to concentrate on a specific code fragment at a time. Meanwhile, workers were collectively contributing to locate the fault related to the same failure. For instance, while one worker was asked whether a certain method call was related to a null pointer exception failure, another worker was asked whether a variable was related to that same failure.

However, the use of questions to partition fault localization into microtasks depended on how we selected the code fragments. After a few feasibility studies we learned that code fragments had to be small, but still contain sufficient information (i.e., lines of code) for workers to answer questions with some degree of confidence. Moreover, since faults could be located anywhere in a Java method, the list of types of the code fragments should cover all lines of code of that Java method.

Following these constraints, we designed the questions around four types of code elements: method calls (program statements), loops (blocks), conditionals (branches), and variables (def-use dependencies). While these code elements tend to be related to small code fragments, these elements correspond to both control and data flow code, which helps to cover the entire source code of a Java method.

#### 3.2.2 Software Bugs used in the Experiment

We searched for a sample set of software bugs by following four criteria. The first criterion was that bugs should be from popular open source projects, so the source code and bug reports would be easily available for someone else to replicate our findings. The second criterion was that the faults should be located in a single Java method, so we could first deal with intra-procedural bugs before dealing with the more complex inter-procedural ones. The third criterion was that the set of bugs should comprise faults in different locations, i.e., in all four code elements and in

Java methods with different sizes. This way, we could evaluate how fault localization outcomes are affected different types of bugs, code elements, source code sizes. For similar reasons, the fourth criterion included bugs that are usually more difficult to locate by only inspecting the source code, for instance, null pointer dereferencing and missing code.

In the search process, we investigated open source repositories, such as Eclipse Bugzilla, GitHub, Apache, and Defect4J. We decided to utilize bug reports from Defect4J [24] because they already contained the information we needed to setup the experiment, i.e., the failure messages, the tests demonstrating each failure and corresponding fix, and the versions of the faulty and fixed source code.

We selected eight bugs from four popular open source projects available in Defect4J. As Table 1 shows, bugs correspond to different failures caused by faults located in all of the four types of code elements pertaining to Java methods with different sizes (lines of code - LOC).

Table 1. Selected bugs (failing Java methods)*

| Java method | project | LOC | failure | fault | related code element |
|---|---|---|---|---|---|
| J1 | Joda Time | 23 | wrong range | missing validation | conditional variable |
| J2 | JFree Chart | 7 | wrong range | wrong variable | variable |
| J3 | Commons Lang | 23 | exception raised | wrong variable | loop conditional variable |
| J4 | JFree Chart | 78 | wrong output | wrong field | conditional method call |
| J5 | Commons Lang | 7 | exception raised | wrong type casting | conditional variable |
| J6 | Closure Compiler | 28 | wrong output | missing conditional | conditional method call |
| J7 | Commons Lang | 12 | exception raised | null pointer dereferencing | loop method call variable |
| J8 | Commons Lang | 33 | wrong format | missing conditional | conditional variable |

*The appendix of this paper contains more detailed information about each Java method.

In order for our experiment to be replicable by others, we included the source code of the faulty Java methods in the appendix. The fixed version of these Java methods can be obtained by querying the Defect4J repository by the project name and bug ID provided in Table 1.

### 3.3 Template Questions

We adopted templates to automatically generate questions about code elements. This way, we avoided manual creation of questions, thereby mitigating the risk of errors or bias from human intervention. Moreover, automatic question generation would make it easier for others to reproduce our experiment.

With regards to questions about source code, our goal was to motivate workers to draw hypotheses about the cause of a failure. Hence, the template questions focused the workers' attention on a possible relation between the source code and the software failure. To do so, we adopted words that conveyed a sense of possibility, for instance, "is there any issue" or "might be related to the failure" (Table 2). This is in line with experiments on online communities, which showed that the use of words reflecting mental processes increased the likelihood of obtaining high-quality responses [71].

Table 2. Template questions

| code element | template question |
|---|---|
| loop | Is there any issue with the loop between lines $x$ and $y$ that might be related to the failure? |
| conditional | Is there any issue with the conditional between lines $x$ and $y$ that might be related to the failure? |
| method call | Is there any issue with the method invocation $M$ at line $x$ that might be related to the failure? |
| variable | Is there any issue with the definition or the use of variable $M$ that might be related to the failure? |

We generated questions automatically by parsing the source code of each Java method and instantiating a question for every occurrence of any of the four types of code elements (Table 2). For our selected set of bugs (Table 1), we instantiated 129 questions (Table 3).

In order to have a ground truth to evaluate whether workers answered the questions correctly, we determined which questions covered source code lines that contained the faults. This was done by comparing the failing and the fixed versions of each Java method.

However, since the fix locations might not necessarily match the root-cause of the failure [72], [73], we inspected the code and executed the unit tests to understand the root-cause of each failure. For the failures related to missing code, the fault location and the fix were not at the same lines. Instead, we considered the fault location to be at the line of code that exposed the failure, e.g., a throwing exception statement, a null pointer dereferencing, or invalid type casting.

Table 3. Questions generated for each Java method

| Questions | J1 | J2 | J3 | J4 | J5 | J6 | J7 | J8 | ∑ |
|---|---|---|---|---|---|---|---|---|---|
| # Instantiated questions | 10 | 6 | 17 | 37 | 9 | 18 | 8 | 24 | **129** |
| #questions covering faults | 2 | 2 | 4 | 5 | 3 | 3 | 3 | 2 | **25** |

### 3.4 TASK DISTRIBUTION AND EXECUTION

We distributed tasks directly on MTurk and made all HITs available to any MTurk worker. Workers could complete HITs independently of each other, thereby enacting a simultaneous workflow [74]. However, each worker could complete up to eight HITs, each HIT pertaining to a different Java method and corresponding fault. In this manner, no single worker could dominate the overall results.

We published all eight HITs on MTurk on Monday at 5 am PCT, so as to maximize the first working hours in the continental US. While our choice of US was because MTurk stopped accepting international workers [41], our choice of time was informed by research that reported the impact of posting time on recruiting [75] and task completion [76].

We chose to distribute microtasks through a type of HIT that allowed microtasks to be performed on an external site, but that required a completion code as evidence that the worker finished the task. In order to perform each HIT, workers had to follow four steps on our external site (Figure 1). First, workers read the study description and digitally signed the consent. Second, workers answered several demographic questions about their age, gender, years of programming experience, programming language, profession (i.e., hobbyist, professional developer, undergraduate student, graduate student), and where the worker learned to program (i.e., high school, university, books, the web). Third, workers took the qualification test. Fourth, workers answered three different questions about the same failing Java method. After this, workers were asked to write a feedback and were finally given the code completion to enter in MTurk.

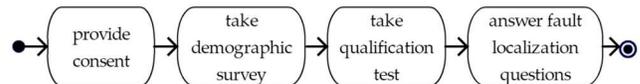

Figure 1. HIT execution steps

In order to answer the questions, workers utilized a web interface containing simple inputs and bug related information displayed on a single page (Figure 2). The interface presented the workers with the following information: (1) unit test that exposed the failure, (2) description of the failure, (3) one code fragment question with three answer options, (4) worker's confidence level on the answer, (5) textual explanation for the answer, and (6) the source code of the Java method. The interface also provided the source code of the methods that invoked or were invoked by the Java method (blue highlight in Figure 2). This provided the worker with some inter-procedural context, which is necessary for answering method call questions. Additionally, after submitting each question, the worker was also provided with options to rank the difficulty level of the question.

We also contemplated a few usability features to help workers tackle the task faster, which is a common concern among workers in microtask platforms. Examples of features included referencing the source code by the corresponding line numbers and highlighting the actual source code (e.g., lines 273 to 275 in Figure 2). Highlighting was also used to help the worker quickly identify the inter-procedural dependencies (as in line 294). We applied a brighter surrounding color for the two regions that we expected the worker to rely more heavily on. We provided a progress bar on the top of the page, so workers could pace themselves by knowing the number of microtasks that remained in the HIT.

Regarding the information that we set to collect, we had some specific design rationales as well. For example, the worker's perceived difficulty and confidence were collected to be later compared with the frequency of correct answers (accuracy). We required textual explanations in order to encourage workers to reflect upon their answers and thereby increase the chances of better quality answers [77], [78]. In the event of workers quitting, we collected their reasons (i.e., "too long", "too difficult", "too boring", "other"), so we could analyze the frequency of these reasons across different workers' profiles and Java method sizes.

![Figure 2 GUI]

Figure 2. GUI for the fault localization microtasks

### 3.5 TASK REPLICATION

We adopted task replication to investigate how many answers from different workers would be necessary to correctly locate the faults. Our underlying assumption was that multiple answers to the same question could be aggregated in ways that would allow us to correctly locate the fault. We will discuss these aggregation mechanisms in section 5.1.

In our experiment, each question was answered by 20 different workers. While this replication level was adopted in other crowdsourcing studies [79], [80], [81], our feasibility studies also utilized a similar replication level of 17. Therefore, we considered 20 answers per question to be a conservative first step.

Regarding the choice of questions to be part of each HIT, questions were randomly selected but followed two constraints. First, all HITs had distinct sets of questions, i.e., each question participated in only one HIT. Thereby, we could distribute HITs through a round-robin approach [82], [83], which would increase the chances that all questions would be answered once before receiving further answers. This would allow us to collect answers uniformly over time, which would help us to analyze how quick the faults were located (or not). Second, questions pertaining to the same HIT should cover code elements that were not adjacent to each other in the source code. This way, we attempted to avoid the situation in which one worker would answer questions about the same few lines of code, which would make the corresponding answers redundant.

## 4 COLLECTED DATA

Before we present the overall results in Section 5, this section presents some of the basic data concerning the number of participants at various stages of the experiment, as well as the demographics of these participants.

### 4.1 Task Participation

The experiment lasted a little less than seven days or, more precisely, 154 hours. We obtained 75% of the answers in the first 48 hours, which is equivalent to a new answer every 40 seconds. Individually, workers took an average of 367 seconds (5 minutes) to submit each answer. The daily peak of activity happened in the mornings (PCT) and the maximum number of concurrent workers was 155.

The final 25% of answers took another five days to collect. This was because it was more difficult to recruit workers towards the end of the experiment and convert recruitment into completed tasks (uptake rate). The reason for this was that, as HITs age in MTurk or simply do not have many remaining available tasks, these HITs become more difficult to be found by potential workers.

Another issue we faced was the rate of worker dropout, as can be seen in the decreasing number of active workers at each step of the process (Figure 3). Out of 3,000 workers who consented to the study, 497 completed at least one HIT. This represents an uptake rate of 16%, similar to a previous MTurk study [84]. The qualification test was responsible for the largest workers' dropout: 1,157 workers quit during the test and 1,050 (60% of those who took the test) did not reach the minimal score to qualify. Hence, out of 1698 workers, only 648 successfully passed the test.

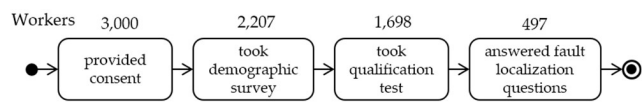

Figure 3. Crowdsourcing workflow with the active workers at each step

Figure 4 shows the scores of all workers who attempted the test. It is worth noting that the workers with a 100% score corresponded to the largest group (40%) of qualified workers.

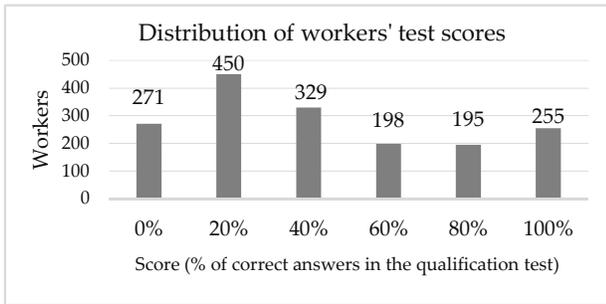

Figure 4. Score distribution of the 1,698 workers who took the test

A second dropout factor consisted of workers quitting in the middle of the microtasks. This is evidenced by the difference between the number of workers who passed the test (648) and the number of workers who completed at least one HIT (497). Out of the 106 workers who provided reasons for quitting, 69 considered the task "too difficult", 11 "too long", 7 "too boring", and 19 "other". The fact that most considered it "too difficult" possibly helped to further filter out lower-skilled workers who nonetheless passed the test.

Since the top reasons for quitting were task difficulty and size (too long), we plotted a chart (Figure 5) to investigate the relationship between the proportion of workers who quit due to difficulty and the size of the Java method. The chart suggests a correlation, which we confirmed to be very strong (Kendall-tau=0.83, z=2.86, p-value=0.004). This implies that larger source code was more difficult, even though each question focused the worker on a small portion of the code.

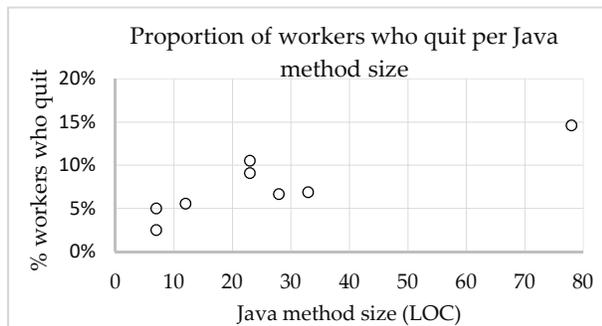

Figure 5. Quits per method size (each dot is a Java method)

Besides worker dropout, the rejection of submissions also reduced the number of workers. The main rejection reason was workers providing invalid completion codes (73 HITs rejected from 52 workers). Interestingly, 42% of these rejected HITs came from workers who had other HITs approved. Many such workers reported reasons related to connection problems, changing computers, or HIT timeout (set to 2 hours). In such cases, we asked workers to retake a different task.

Overall, the average number of HITs per worker was 2.3 and the median was 1. The distribution (Figure 6) shows that a few workers did not dominate the results.

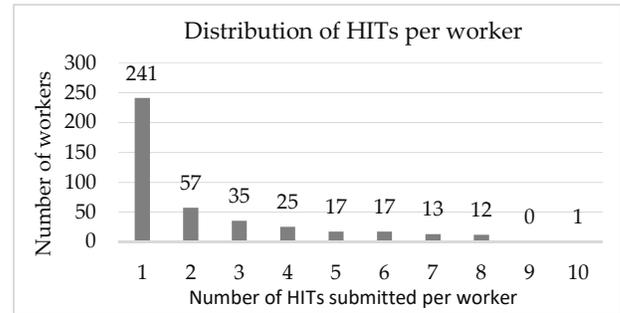

Figure 6. Distribution of HITs submitted per worker

## 4.2 Worker Demographics

Different studies report quite stark differences in MTurk demographics over the years [85], [86], [87]. Reasons appear related to be the fact that different tasks attract different people [88], as well as changes in MTurk policies [41]. Besides this inherent uncertainty about the type of workers we would attract, we were unsure about how accurate the self-reported data would be. Hence, we looked for the consistencies between the demographics presented in Table 4 and the demographics in the recent literature.

Table 4. Demographics of the 2,207 respondents

| | |
|---|---|
| **Gender** | Males (65%), Females (34%), Others (1%) |
| **Country of residence** | USA (71%), India (17%), Others (12%) |
| **Profession** | Graduate students (16%), Undergraduate students (25%), Professional developers (20%), Hobbyists (30%), Others (9%) |
| **Top programming languages** | Java (25%), C/C++/C# (20%), Python (7%) |
| **Learned to code at** | University (44%), The Web (30%), High school (19%), Others (7%) |

The gender distribution in Table 4 suggests that our study attracted more female workers (34%) than the usual proportion of females in computer programming activities according to Taulbee [89] and NSF [90] reports. Interestingly, among professionals, the percentage was somewhat lower, namely 26%.

The composition of 71% of US workers is consistent with the restrictions MTurk imposed on international workers [41]. This is also in line with reported figures for MTurk workers [87].

Regarding test scores, 57% of professional developers passed the skill test, whereas only 35% who declared themselves as non-professionals (students, hobbyists, and others) passed the test. Professional developers who passed the test were also less likely to quit, while hobbyists and undergraduates were the two groups who most frequently quit, corresponding to 63% of workers who quit. These results are not too surprising, except for perhaps the fact that only 57% of professionals passed the qualification test. We have no good explanation for why this was not higher, other than perhaps random interest and the professionals not taking the test seriously.

The self-reported number of years of programming experience (YoE) was also consistent with professions and test scores. Professional developers were on average significantly more experienced than the other workers (Wilcoxon test, w = 611628, p-value <0.0001). Meanwhile, students were the least experienced and showed the smallest distribution spread (Figure 7). Regarding the test scores, workers who passed the test reported 7.8 YoE, while workers who did not pass reported 3.7 YoE. This difference was statistically significant (Wilcoxon test, w= 477222, p-value <0.0001). It is also worth noting that the YoE for professional developers and graduate students are like the ones reported in recent research that discusses the use of students in software engineering experiments [91]

Finally, concerning data on programming knowledge. The three most popular programming languages reported (Table 4) also ranked similarly high in a recent IEEE survey [92]. Regarding where workers learned how to code, the majority (63%) reported a traditional educational setting (university and/or high school). Although the Web can be a source of formal and informal training, 60% of workers who selected the Web also selected university or high school (or both).

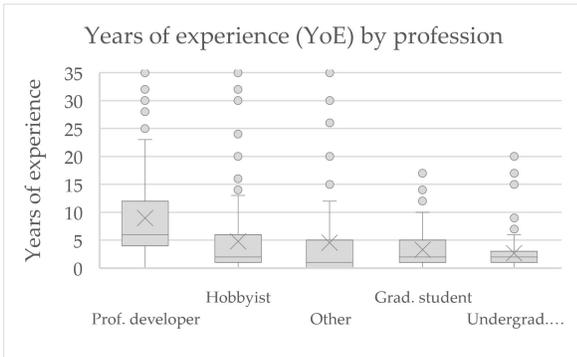

Figure 7. Years of programming experience, not showing outlier [prof. "Other", 50 YoE] corresponding to a 72 years old worker

## 5 DATA ANALYSIS

To determine conditions in which the crowd could locate a fault, we examined three different aggregation mechanisms (AM) in terms of effectiveness, cost, and speed.

### 5.1 Aggregation Mechanisms

A challenge in any microtask-based crowdsourcing consists of how to take individual answers and turn them into a "collective wisdom" [93]. Different aggregation approaches have tackled this challenge [80], [94], [5], [96], [81], [97], [98]. In our case, we had to decide how to detect which answers for a question should be aggregated into a single prediction. For this, we explored three different aggregation mechanisms.

- AM.1: If a question receives $n$ more YES answers than NO answers, the crowd predicts that there is a fault in the source code covered by the question. This is a relative aggregation mechanism, which is based on the presumption that most workers can identify a fault when it is there.
- AM.2: If a question receives more YES answers than a certain number $n$, the crowd predicts a fault is located in the source code covered by the question. This is an absolute aggregation mechanism, which simply uses a threshold to decide whether a crowd predicts a fault.
- AM.3: If a question is among the top $n$ questions which received the largest number of YES answers, the crowd predicts that a fault is located in the source code covered by the question. Instead of deciding question by question, the aggregation mechanism examines all questions and takes the top $n$ as predictive of where the fault is.

We call the first two aggregation mechanisms "within question" because each question is treated independently. We term the third aggregation mechanism "across-questions", as it compares the judgments of different groups of workers, each group having answered a different question.

To apply these aggregation mechanisms, we needed to determine the values of $n$ for each of them. Since our experiment involved asking each question 20 times, the range of values for $n$ was from zero to 20. For each value of $n$, we applied the aggregation mechanism to determine the questions that the crowd considered as the ones covering faults. Since we knew which questions covered faults, we could identify two groups of questions: questions that were correctly considered to cover faults (i.e., true positives) and questions correctly considered to not cover faults (i.e., true negatives).

The results of the first aggregation mechanism (Figure 8) show that, while more faults would be correctly located (true positives) using smaller values of $n$, there would be fewer correct predictions about questions that did not cover the faults (true negatives). This means that as more faults were located, more of the questions that did not cover faults would be incorrectly considered to be ones covering faults (false positives). This trade-off is better visible in the outcomes of AM.2 (Figure.9) and AM.3 (Figure.10).

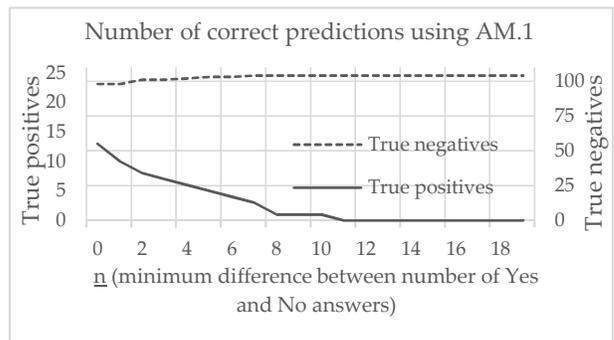

Figure 8. Correct predictions by considering different values of $n$ (minimum difference between number of YES and NO answers to consider a question as fault covering).

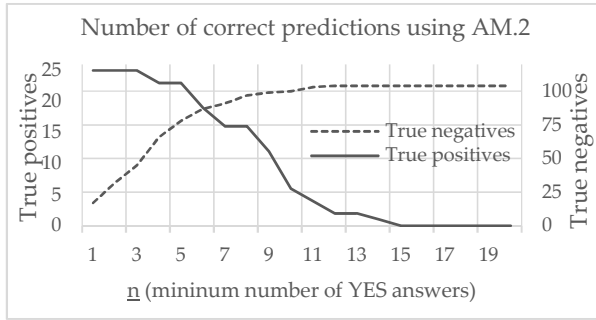

Figure 9. Correct predictions at each level of n (minimum number of YES answers to consider each question as covering a fault).

In order to understand situations in which the crowd would have produced the best result, we chose a value of n through which the aggregation methods located all faults and generated the lowest number of false positives. In the context of our experiment, the values of n correspond to: n=0 for AM.1, n=5 for AM.2, and n=2 for AM.3.

Note that these values of n do not correspond to the maximum number of true positives in Figure 8, Figure 9, or Figure 10. The reason for this is that questions overlap in the lines of code they covered. Therefore, for faults at a single line of code, one true positive outcome per Java method is enough to correctly locate the fault.

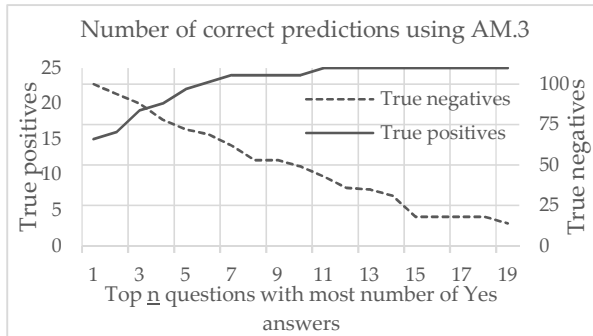

Figure 10. Correct predictions by considering different top n questions with most number of YES answers.

Table 5 shows the outcomes of the aggregation mechanisms for each of these values of n. While the crowd would have correctly predicted all faults using AM.2 and AM.3, the crowd would have failed to predict one of the faults using AM.1. Since these are simply the sum of correct predictions, in the next section investigate the effectiveness of predictions at the method level and line-of-code level.

TABLE 5. FAULTS LOCATED

|  | AM1. | AM2. | AM3. |
|---|---|---|---|
| true positives | 13 | 19 | 15 |
| false positives | 6 | 17 | 4 |
| false negatives | 12 | 6 | 10 |
| true negatives | 98 | 87 | 100 |
| total | 129 | 129 | 129 |
| faults located | 7 | 8 | 8 |

## 5.2 Effectiveness

We study effectiveness of the crowd at two levels of granularity, at the question level and the line-of-code level. The reason stems from an important characteristic of our approach, and more specifically of the questions that we have included, that some questions overlap. That is, a line of code can be covered by multiple questions. This influences how we decide whether a fault has been identified, i.e., at question or at line-of-code level.

### 5.2.1 Question level

At the question level, we declare that a fault is correctly located if at least one of the questions that cover the faulty line(s) of code is correctly predicted by the aggregation mechanism as the one covering a fault. This can be illustrated by the numbers provided in Table 5. Through AM.2 and AM.3, the crowd would have correctly predicted all of the eight faults even when a few questions were not correctly identified, as shown in the number of false negatives and false positives.

From Table 5 we can also see that, while AM.2 correctly identified more fault covering questions (i.e., fewer false negatives), AM.1 and AM.3 generated fewer false positives. This suggests that a choice of aggregation method would follow a trade-off between either minimizing false negatives or false positives.

We explore this trade-off in more detail by computing the outcome of each aggregation mechanism for each Java method. To consistently compare outcomes from Java methods with different sizes, we utilized relative measures of false positives (i.e., precision) and false negatives (i.e., recall). Results are presented in Figure 11 and Figure 12.

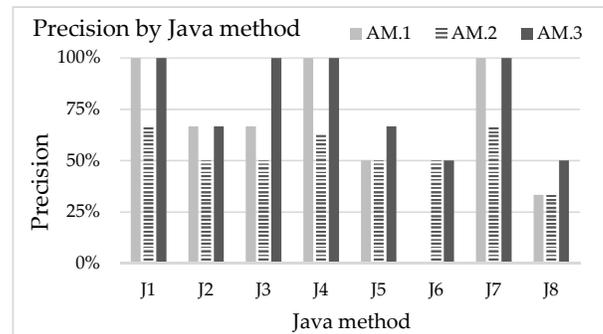

Figure 11. Precision levels by Java method.

The trade-off between false positives and false negatives was also present for most Java methods. Figure 11 shows that, for most Java methods, AM3 presented a lower or equal proportion of false positives (i.e., higher precision), whereas Figure 12 shows that AM.2 and AM.1 presented a lower proportion of false negatives (i.e., higher recall) for most Java methods. Since most of our faults are located at a single line of code and questions overlap in lines of code covered, higher levels of recall do not imply that more faults were located (as far as at least one question is correctly predicted as a true positive).

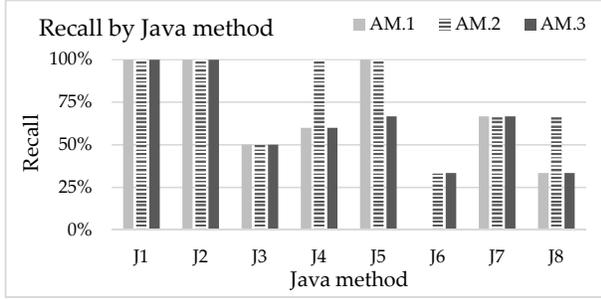

Figure 12. Recall levels by Java method

It is worth noting that the same two Java methods (J6, J8) are among the four Java methods with lowest precision values under AM.3 and lowest recall value under AM.1 and AM.2. These two Java methods (J6, J8) contained faults related to missing source code, which is usually a fault that is more difficult to locate [99], [100]. Also among the lowest precision values are two of the smallest Java methods (J2, J5) in terms of the number of lines of code (LOC). Based on these evidences, we suggest that further studies are necessary to investigate the effect of code size and fault type on programmers inspecting source code for faults.

### 5.2.2 Line of code level

Just knowing whether one or more questions identify a fault does not necessarily give a sense of the amount of effort that a developer might invest in order inspect all the lines of code covered by a question (or multiple questions, if more than one question predicts a fault to be somewhere in the Java method). To address this issue, we performed a second analysis, one in which we focus on how many lines of code a developer must inspect given an aggregation method. In order to discriminate between source code lines that contain faults and those that do not, but which still need to be identified, we created a set of categories.

- True positive line: a line where a fault is correctly located. In our experiment, six of the actual faults are located on a single line and two faults (J4, J5) are located on two lines.
- Near positive line: any line different from the true positive line, but that was covered by the same question for which the crowd correctly considered to cover a fault (true positive).

Based on these categories, we computed the number of lines considered faulty by each of the three aggregation mechanisms (Table 6). AM.3 also presented the fewest false positives at the line-of-code level. It located all ten faulty lines and considered the fewest number of lines as false positives.

TABLE 6. LINES TO INSPECT

| Line categories | AM.1 | AM.2 | AM.3 |
|---|---|---|---|
| true positive lines | 9 | 10 | 10 |
| near positive lines | 29 | 35 | 31 |
| false positive lines | 1 | 9 | 2 |
| false negative lines | 0 | 0 | 0 |
| true negative lines | 144 | 157 | 168 |
| total lines* | 183 | 211 | 211 |
| extra lines to inspect** | 30 | 44 | 33 |
| % of total lines | 16% | 21% | 16% |

\* total source code lines of the Java methods which faults were located
\** near positive plus false positive lines (do not count true positive lines)

While it is an unavoidable effort to inspect the line where the fault is located, inspecting near and false positive lines are extra effort, which we showed in the last two lines of Table 6. However, we show for the outcomes of AM.3 in Table 7, that this effort varies from 1% to 43% of the lines of code across different Java methods. We can also notice in Table 7 that the largest Java method (J4) presented the smallest proportion of lines to inspect (1%) while the smallest Java method (J2) presented the highest proportion of lines to inspect (43%). This suggest that the cost-effectivity of crowdsourcing fault localization might be dependent of the size of the source code involved in the failure. i.e., smaller source codes might be less cost-effective than larger source codes.

TABLE 7. AM.3 OUTCOMES AT LINE OF CODE LEVEL

| lines | J4 | J6 | J7 | J1 | J5 | J3 | J8 | J2 |
|---|---|---|---|---|---|---|---|---|
| true positive lines | 2 | 1 | 1 | 1 | 2 | 1 | 1 | 1 |
| near positive lines | 1 | 0 | 1 | 5 | 2 | 8 | 12 | 2 |
| false positive lines | 0 | 1 | 0 | 0 | 1 | 0 | 0 | 1 |
| false negative lines | 0 | 0 | 0 | 0 | 0 | 0 | 0 | 0 |
| true negative lines | 75 | 26 | 10 | 17 | 3 | 14 | 20 | 3 |
| total lines* | 78 | 28 | 12 | 23 | 7 | 23 | 33 | 7 |
| extra lines to inspect** | 1 | 1 | 1 | 5 | 2 | 8 | 12 | 3 |
| % total lines | 1% | 4% | 8% | 22% | 29% | 35% | 36% | 43% |

\* total source code lines of the Java methods which faults were located
\** near positive plus false positive lines (do not count true positive lines)

Observe, however, that the extra lines to inspect are mostly near positive lines and 80% of these lines are concentrated in three Java methods (J1, J3, J8). By studying the structure of these Java methods (see appendix), we see that two of them (J3, J8) presented the largest proportion of lines covered by multiple questions. These were the cases, for example, in which a method call pertains to a conditional branch that is inside a loop. Since we have one question for each of these items, the lines nested deeper in the source code were covered by multiple questions. Hence, when the fault happens to be located on these deeper lines, the chances of selecting more near positive lines for inspection are higher. A larger sample with Java methods of different sizes and diverse internal structures would help test the statistical significance of these observations.

## 5.3 Cost

Our total cost for the experiment was $3,077.00. Naturally, a question is whether we could possibly have done equally well with a replication factor of less than 20 answers per question. To do this, we calculated for each aggregation mechanism considering one, two, three, and so forth answers per question.

Figure 13 shows for each Java method, how many answers were necessary to locate the corresponding fault. While AM.2 would require a minimum of 20 answers/question to locate all faults (see method J6), AM.1 and AM.3 would require 14 answers (J5). However, note that AM.1 did not locate all faults.

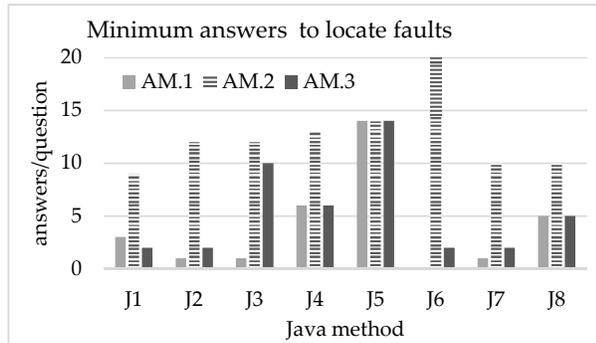

Figure 13. Minimum answers per question to locate all faults

Looking at the lines to inspect at these answer levels, the number of lines to inspect from partial answers (Table 8) is also larger than the number of lines to inspect considering all 20 answers per question (Table 6). These differences are important because they suggest a trade-off between answers per question (cost) and lines to inspect (effectiveness).

TABLE 8. TOTAL LINES TO INSPECT CONSIDERING THE MINIMUM LEVEL OF ANSWERS FOR EACH JAVA METHOD FAULT

| line categories | AM.1 | AM.2 | AM.3 |
|---|---|---|---|
| true positive lines | 9 | 10 | 10 |
| near positive lines | 22 | 30 | 32 |
| false positive lines | 21 | 2 | 18 |
| total lines | 183 | 211 | 211 |
| extra lines to inspect* | 43 | 32 | 50 |
| % of total lines | 23% | 15% | 23% |

* near positive plus false positive lines

## 5.4 Speed

We are interested in analyzing how long it would take to locate all faults if we could rely on partial sets of answers. The partial sets were obtained by finding cut times at which all questions had the same number of answers. Table 10 shows the results after utilizing AM.3, which we chose because it consistently located faults after a certain number of answers. This was not true for AM.1 and AM.2 (see appendix) the best results in terms of requiring fewer answer per question to locate all the eight faults. The outcomes from AM.1 and AM.2 are available in the appendix (Table 9)

TABLE 10. SPEED TO LOCATE FAULTS

| cut time (hours) | A* | workers | answers | faults located | lines to inspect |
|---|---|---|---|---|---|
| 7.2 | 1 | 35 | 129 | 7 | 37 |
| 10.1 | 2 | 66 | 258 | 8 | 73 |
| 12.8 | 3 | 96 | 387 | 7 | 42 |
| 21.5 | 4 | 128 | 516 | 7 | 38 |
| 27.3 | 5 | 158 | 645 | 8 | 46 |
| 31.4 | 6 | 181 | 774 | 7 | 52 |
| 34.4 | 7 | 210 | 903 | 7 | 48 |
| 40.2 | 8 | 243 | 1032 | 7 | 40 |
| 76.4 | 9 | 267 | 1161 | 7 | 35 |
| 77.5 | 10 | 288 | 1290 | 8 | 43 |
| 78.1 | 11 | 309 | 1419 | 8 | 53 |
| 85 | 12 | 339 | 1548 | 8 | 52 |
| 98 | 13 | 363 | 1677 | 8 | 52 |
| 108.1 | 14 | 386 | 1806 | 8 | 53 |
| 108.6 | 15 | 417 | 1935 | 8 | 55 |
| 123.5 | 16 | 446 | 2064 | 8 | 55 |
| 131.3 | 17 | 460 | 2193 | 8 | 54 |
| 138 | 18 | 480 | 2322 | 8 | 53 |
| 149.7 | 19 | 490 | 2451 | 8 | 53 |
| 154.7 | 20 | 497 | 2580 | 8 | 43 |

*A = answers per question

It is worth noting that the time intervals between cut times varied and increased towards the latter hours of the experiment. We confirmed this with a linear regression of the time intervals, which showed an upward trend line with intercept at 3 hours and alpha of 0.32. This was obtained after removing one outlier corresponding to the time interval between the cut times 40.2 and 76.4 hours (Table 10). At the same time, the number of workers recruited also decreased. This quantifies what we highlighted in Section 4, that we collected 75% of the answers in the first 48 hours. Hence, the overall speed of completing microtasks decreased as the time passed, which hindered the completion of the remaining microtasks. This speed variation is also visible in the horizontal distances of consecutive data points in Figure 14.

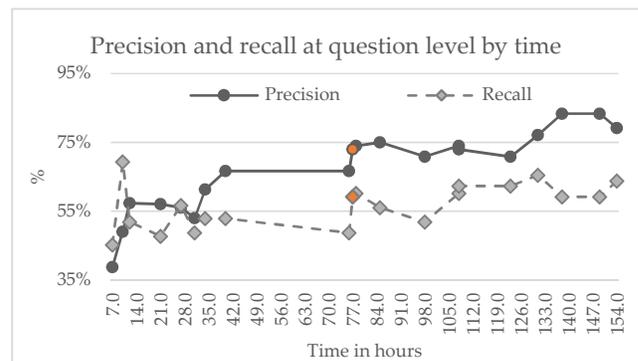

Figure 14. Precision and recall at each cut time (red dot marks the time when all faults were located, i.e., 77.5 hours)

As the speed of collecting answers decreased towards the end of the experiment, the variation in the number of lines to inspect also decreased (Table 10). This is a positive factor because it suggests that faults are located before the completion of all microtasks. Although all the 8 faults were located at cuts times 10.1 and 27.3 hours, only after 77.5hours, all faults were located regardless of the additional answers collected. This is also visible in the recall curve in Figure 14 and Figure 15. After a certain level (50% recall), additional answers do not necessarily increase the number of faults located.

Concerning the lines to inspect at each cut time, Figure 15 shows that as more answers were collected, the number of false positives dropped. However, the number of near positives increased, which means that more overlapping questions were the ones covering faults. This suggests that, due to the existence of overlapping questions, collecting more answers above a certain threshold might not increase the precision of fault localization at the line level. This is in sharp contrast to Figure 14, which shows an increase in precision at the question level.

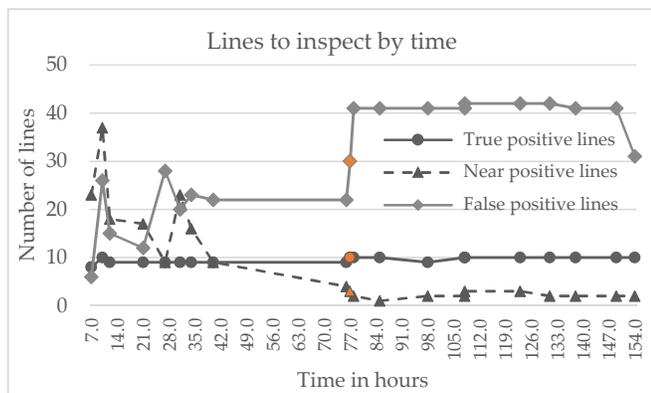

Figure 15 Lines to inspect at each cut time (red dot shows the time when all faults were located, i.e., 77.5h)

## 5.5 Filtering answers

Although the crowd of programmers located all faults within a few lines of code and by relying on fewer than 20 answers per question, we wondered if we could have done better than that. In order to explore this, we investigate whether subgroups of workers (i.e., subcrowds) had also located the eight faults. Whenever this was true, we looked at the number of workers, answers, and lines they flagged for inspection. We selected these subcrowds by composing filters based on the attributes of questions, answers, and workers. Our objective was not to build predictors of the best subcrowd to locate a fault. Instead, our goal was to explore the data in a structured way by building on intuition and insight.

### 5.5.1.1 Could we locate all faults by asking questions according to lines covered?

The intuition is that questions covering more lines of code would have a lower accuracy of answers. We explored this intuition for the variable and conditional question types. The reason is that, the loop question type had only four concrete questions and the method call questions covered a single line of code each.

Restricted to variable and conditional question types, we studied the correlation between lines of code covered by each question and the accuracy of answers to these questions. While variable question type did not show statistical significant correlation (p-value = 0.2882), conditional type question showed significant negative correlation ($z = -2.0219$, p-value = 0.04318, Kendall tau = -0.2575). The negative correlation suggests that answers were less accurate for questions covering more lines of code.

Based on this evidence, we investigated filters that prioritized questions covering fewer lines of code. We learned that if we filter out answers from conditional questions that cover more than 3 LOC, the corresponding subcrowd will have located all eight faults (Table 11). This suggests that prioritizing questions covering fewer lines of code, particularly questions about conditionals, might be an effective heuristic to locate faults.

TABLE 11. FILTERING BY CONDITIONAL QUESTIONS THAT COVERED MORE THAN THREE LINES OF CODE

| Filtered out answers | precision, recall | lines to inspect | workers | answers |
|---|---|---|---|---|
| conditional covering > 3 LOC | 77%, 76% | 34 | 494 | 2280 |

### 5.5.1.2 Could we locate all faults by selecting questions by difficulty?

Our intuition is that questions that are considered more difficult would be less accurate. Corroborating this intuition, a previous human study [100] showed that programmers perceived conditional branches as more difficult to debug. Therefore, answers from questions about these type of code elements would be less accurate.

We investigated differences in workers' perceived difficulty among question types of the same Java method and across all Java methods. The Wilcoxon rank test did not show any statistically significant differences among question types in terms of difficulty level. Hence, the answer to the exploratory question is that we could not use difficulty as a parameter to select questions.

One possible reason for the lack of evidence could be that the workers' perception of difficulty was based on the Java method and not on the question and respective code element. In order to investigate this, we examined whether Java methods were significantly distinct in terms of the average perceived difficulty of the questions asked about them. We performed a Wilcoxon rank test with a Bonferroni correction for multiple two-by-two comparisons (adjusted p-value =0.0018). The test showed that the differences in difficulty level were statistically significant for only 10 out of the 28 pairs of Java methods.

Hence, other confounding factors besides Java method complexity and size could have affected workers' perceived difficulty. Further experiments would be necessary to understand what affects a worker's perceived difficulty during fault localization tasks.

### 5.5.2 Filters based on answers

The goal of this section is to investigate if answer attributes such as confidence, difficulty, duration, and explanation size could be used to filter out low accurate answers.

#### 5.5.2.1 Could we locate all faults by filtering out answers according to workers' confidence and difficulty?

One would expect that workers were more confident about their answers when they find the question less difficult. We investigated this intuition by computing the correlations between confidence and difficulty for each Java method (Table 12) The medium and strong negative correlations confirm this intuition.

TABLE 12. CONFIDENCE AND DIFFICULTY CORRELATIONS*

|  | J1 | J2 | J3 | J4 | J5 | J6 | J7 | J8 |
|---|---|---|---|---|---|---|---|---|
| Correlations | -0.56 | -0.37 | -0.34 | -0.42 | -0.49 | -0.46 | -0.55 | -0.53 |

*Significant Kendall-tau correlations, p-value < 0.05

These correlations also suggest that answers might be concentrated on certain pairs of difficulty and confidence values, for instance, pairs such as low confidence and high difficulty or high confidence and low difficulty. In order to investigate this, we displayed the number of answers categorized under each difficulty/confidence value pairs, which are represented by 30 cells in Table 13.

Note that, if workers had randomly categorized their answers in terms of difficulty and confidence, the probability of an answer pertaining to any cell would be 3.33% (1/30). However, Table 13 shows that the answers were concentrated in certain cells. We showed that by highlighting cells containing more answers than expected from a random distribution (i.e., 3.33% x 2580 = 86 answers). This criterion has been adopted in other crowdsourcing studies to identify whether a crowd reached a consensus [101]. Note that this does not mean that the consensus is correct.

TABLE 13. ANSWERS BY PAIRS OF DIFFICULTY/CONFIDENCE (HIGHLIGHTED CELLS HAVE MORE THAN 3.3% OF ANSWERS)

|  |  | Difficulty | | | | |
|---|---|---|---|---|---|---|
|  |  | 1-low | 2 | 3 | 4 | 5-high |
| Confidence | 0-IDK | 2 | 12 | 46 | 90 | 156 |
|  | 1-low | 8 | 5 | 10 | 20 | 40 |
|  | 2 | 3 | 20 | 43 | 79 | 32 |
|  | 3 | 2 | 27 | 276 | 150 | 48 |
|  | 4 | 31 | 218 | 254 | 163 | 33 |
|  | 5-high | 355 | 207 | 126 | 57 | 67 |

* 0-IDK values are automatically set when a worker selects IDK.

The fact that answers are concentrated in certain cells is useful only if the answers in these cells are more accurate than answers in the adjacent cells, which is not the case as Table 14 shows. Hence, the only aspect that made these cells stand out was that the crowd answers were concentrated in these pairs of difficulty/confidence.

TABLE 14. AVERAGE ACCURACY OF DIFFICULTY/CONFIDENCE

|  |  | Difficulty | | | | |
|---|---|---|---|---|---|---|
|  |  | 1-low | 2 | 3 | 4 | 5-high |
| Confidence | 0-IDK | - | - | - | - | - |
|  | 1-low | 100% | 85% | 67% | 61% | 50% |
|  | 2 | 50% | 59% | 67% | 67% | 56% |
|  | 3 | 84% | 83% | 62% | 63% | 48% |
|  | 4 | 84% | 75% | 77% | 72% | 64% |
|  | 5-high | 100% | 100% | 90% | 50% | 73% |

After filtering out the answers outside to these highlighted cells, the corresponding subcrowd would have located all eight faults (Table 15).

TABLE 15. FILTERING BY DIFFICULTY/CONFIDENCE PAIRS

| Filtered out answers | precision, recall | lines to inspect | workers | answers |
|---|---|---|---|---|
| Pairs with less than 3.3% of the answers | 71%, 56% | 42 | 412 | 1749 |

### 5.5.3 Could we locate all faults by filtering answers by duration?

Workers took significantly more time to answer the first question than the second and third questions, which is visible in Figure 16. We also confirmed that the differences among the average answer durations of these three group of answers are statistically significant (Wilcoxon test, p-value<0.05).

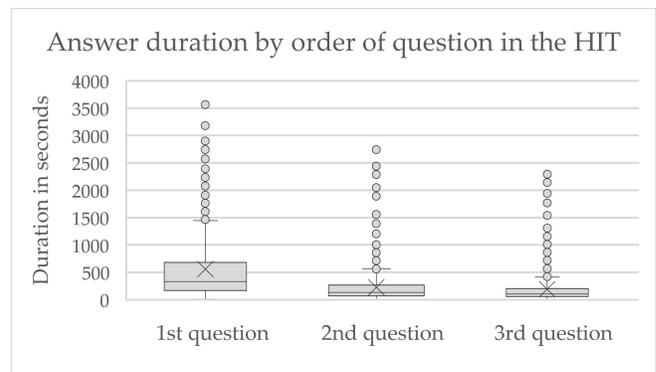

Figure 16. Duration of answer by order of question answered in by each worker in each HIT

In order to study the relation between duration and proportion of correct answers, we grouped answers by the distribution quartiles and by the order of the answer (Table 16). We noticed a few patterns. The first quartile has the lowest proportion of correct answers among the four quartiles. The proportion of correct answers increases from left to right. Moreover, for all three questions, there is an increase of accuracy answers from the first to the third quartile.

TABLE 16. ACCURACY OF ANSWERS BY QUARTILE OF DURATION

| Duration quartiles | % correct answers by order in HIT | | |
|---|---|---|---|
| | 1st question | 2nd question | 3rd question |
| 1st QT | 61% | 69% | 75% |
| 2nd QT | 68% | 74% | 77% |
| 3rd QT | 68% | 77% | 80% |
| 4th QT | 64% | 69% | 76% |

Based on these patters we tried different duration filters. All faults were located only when we filtered out the fastest answers (first quartile). Table 17 shows the result.

TABLE 17. FILTERING THE FASTEST ANSWERS

| Filtered out answers | precision, recall | lines to inspect | workers | answers |
|---|---|---|---|---|
| 1st Quartile duration | 78%, 59% | 41 | 488 | 2395 |

#### 5.5.3.1 Could we locate all faults by filtering answers by explanation size?

Drawing inspiration from research that shows that explanations are related to better quality answers [102], [78], we investigated if a longer textual explanation could relate to more accurate answers. We did this regardless of answer option because textual explanations sizes were not statistically distinct among answer options (Figure 17).

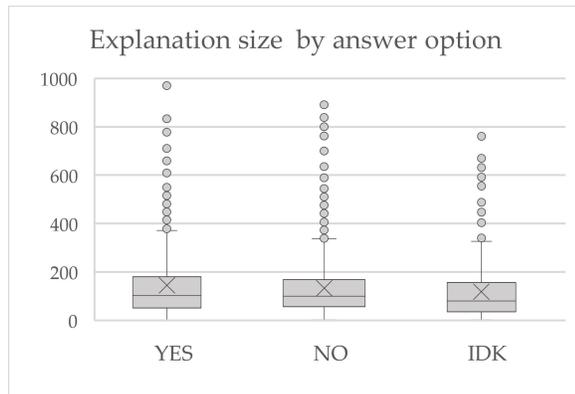

Figure 17. Explanation sizes (not showing outliers – explanation sizes above 1000 characters)

We noticed that only the answers from the lowest quartile had accuracy value (56%) smaller than the answers in the 2nd, 3rd, and 4th quartiles (respectively 65%, 66%, and 64% proportion of correct answers). This pattern is also true at the Java method level, as Table 18 shows that for most methods, the shortest explanations (1st quartile) also came from answers that are more frequently incorrect.

At the Java method level, longer explanations are also related to more accurate answers. The last column in Table 18 shows that for most Java methods, answers pertaining to the 1st quartile have the lowest accuracy values.

TABLE 18. ACCURACY BY QUARTILE OF EXPLANATION SIZE

| Explanation quartiles | J2 | J3 | J5 | J7 | J4 | J1 | J6 | J8 | ∑* |
|---|---|---|---|---|---|---|---|---|---|
| 1st QT | 41% | 50% | 50% | 54% | 55% | 57% | 59% | 68% | - |
| 2nd QT | 68% | 50% | 56% | 67% | 60% | 85% | 60% | 82% | 7 |
| 3rd QT | 50% | 62% | 45% | 63% | 64% | 93% | 59% | 74% | 6 |
| 4th QT | 67% | 59% | 56% | 61% | 66% | 70% | 63% | 66% | 7 |

*∑ = counts how many methods have quartile accuracy > first quartile accuracy

Therefore, it would make sense to filter out answers from the 1st quartile. After doing so, the corresponding subcrowd located all eight faults with the following statistics (Table 19).

TABLE 19. FILTERING THE SHORTEST EXPLANATIONS

| Filtered out answers | precision, recall | lines to inspect | workers | answers |
|---|---|---|---|---|
| 1st Quartile size of explanation | 70%, 66% | 57 | 412 | 1879 |

### 5.5.4 Filters based on worker attributes

We selected four worker attributes: years of programming experience (YoE), profession, score in the qualification test, and proportion of IDK answers. In order to have a fair basis to compare workers with each other, we computed correlations and differences by considering only the outcome of the first HIT of each worker (i.e., first three questions).

#### 5.5.4.1 Could we locate all faults by filtering out workers according to years of experience (YoE)?

We expected that more experienced workers would have a higher accuracy of answers. At the same time, we saw in section 4 (Figure 7) that years of experience have different distributions across professions, with students and others presenting the lowest levels of experience. In order to investigate if YoE is related to accuracy, we computed the correlation between YoE and accuracy of workers' answers (Kendall tau=0.18, p-value<0.0001). While this significant correlation confirms our intuition, the weak value (tau = 0.18) suggests that not all professions and levels of YoE may present better answer accuracy.

After breaking down accuracy by quartiles of YoE, we can see two patterns (highlighted in Table 20): the top quartile presents the highest accuracy values among professions and only for professional developers do higher levels of YoE correspond to increasing levels of accuracy. Nonetheless, after filtering out answers from workers outside these groups, the corresponding subcrowds did not locate all the eight faults. Therefore, the answer for the exploratory question is negative.

TABLE 20. ACCURACY OF ANSWERS BY QUARTILE OF YOE

| Professions | Quartiles of Years of Experience | | | | |
| --- | --- | --- | --- | --- | --- |
| | 1st | 2nd | 3rd | 4th | all |
| Other | 67% | 62% | 64% | 78% | 75% |
| Prof. Developer | 57% | 70% | 75% | 75% | 71% |
| Hobbyist | 63% | 69% | 67% | 75% | 71% |
| Graduate | 61% | 54% | 52% | 82% | 68% |
| Undergraduate | 63% | 56% | 60% | 74% | 66% |

#### 5.5.4.2 Could we locate all faults by filtering out workers according to profession?

In order to investigate this question, we first need to know whether workers grouped by professions are significantly distinct in terms of the average accuracy of their answers. Table 21 shows that this is true for a few professions. Although the effect sizes are small (values between 0.1 and 0.3), this suggests that filtering out workers by profession could lead to better selection of answers. Note that "Undergraduates" and "Others" did not show any significant difference among the other professions.

TABLE 21. DISTINCT PROFESSIONS

| Groups of workers with statistically significant differences | Effect size |
| --- | --- |
| Professional developers x Graduate students | r = 0.1551 |
| Students* x Non-students | r = 0.1323 |
| Professional developers x Hobbyist | r = 0.1193 |

* students=graduate and undergraduate students

After filtering out several combinations of these professions, we found that four subcrowds have located all eight faults. (Table 22).

TABLE 22. FILTER BY WORKER PROFESSION

| Filtered out workers | precision, recall | lines to inspect | workers | answers |
| --- | --- | --- | --- | --- |
| students, hobbyists, others | 65%,58% | 46 | 232 | 1260 |
| hobbyists, graduate students, others | 71%,54% | 37 | 287 | 1513 |
| students, others | 75%,70% | 55 | 280 | 1578 |
| students | 81%,76% | 56 | 316 | 1773 |

#### 5.5.4.3 Could we locate all faults by filtering out workers according to their score in the qualification test?

Workers' qualification test scores were positively correlated with the number of correct answers (Kendall tau=0.13, p-value<0.0001). i.e., the workers with larger qualification test score also were more accurate in their microtasks. Additionally, when we grouped workers by score, a few groups were significantly (p-value<0.05) distinct from each other in terms of the average accuracy of their answers (Table 23).

TABLE 23. DISTINCT WORKERS BY SCORE

| Groups of workers with statistically significant differences | Effect size |
| --- | --- |
| 100% versus 60% score | r = -0.1890 |
| Above 80% versus 60% score | r = -0.1849 |
| 80% versus 60% score | r = -0.1311 |

Note that comparing workers according with their score (Table 23) produced larger effect sizes than comparing workers by profession (Table 21). This suggests that worker score is a better criterion than profession to segregate groups of workers in terms of the average accuracy of their answers.

Therefore, we expected that the outcomes from groups of workers selected by score would be better. After applying filters on worker score, we found two subcrowds would have located all faults (Table 24). Since the outcomes from these subcrowds are better than the ones filtered by profession, we confirmed our initial intuition.

TABLE 24. FILTERS BY WORKER SCORE

| Filtered out workers* | precision, recall | lines to inspect | workers | answers |
| --- | --- | --- | --- | --- |
| scored 60% U 80% | 75%, 57% | 42 | 351 | 567 |
| scored 100% | 94%, 65% | 51 | 194 | 311 |

* U = union of sets

#### 5.5.4.4 Could we locate all faults by filtering out workers who answered IDK more frequently than others?

The intuition is that a large proportion of IDK answers could represent the difficulty faced by certain workers in accomplishing the tasks. Hence, these workers would have higher chances of providing wrong answers. If that is true, we could obtain better answers by filtering out workers based on their proportion of IDK answers. In order to evaluate this intuition, we needed to confirm two assumptions: that IDK answers were related to difficult answers and that workers grouped by the proportion of IDK answers also had a distinct proportion of correct answers.

Concerning difficulty level, more IDK answers were considered difficult than easy. Table 25 shows that 80% of IDK answers were considered to be very difficult (levels 4 or 5). Meanwhile, YES answers were equally distributed among difficulty levels and NO answers were considered easier than IDK answers. Looking at the Java methods individually (Table 26), the IDK answers were more difficult across all methods, which was not the case for the other answer options.

This distribution of difficulty levels was also consistent across workers' professions. Workers who were professional developers contributed to 40% of YES or NO answers, but these workers contributed to only 20% of IDK answers. Hobbyists and undergraduate students contributed to 60% of IDK answers. This suggests that the other worker professions were more frequently clueless than the professional developers.

Likewise, we would expect similar results for workers with a lower score in the qualification test. However, there were no statistical significant differences among the average score of workers who answered IDK versus workers who answered YES or NO.

TABLE 25. PROPORTION OF ANSWERS BY DIFFICULTY LEVEL

|  | % of answers at each level of difficulty* | | | | | |
|---|---|---|---|---|---|---|
|  | 1 | 2 | 3 | 4 | 5 | Total |
| IDK | 1% | 4% | 15% | 29% | 51% | 100% |
| YES | 16% | 20% | 28% | 20% | 15% | 100% |
| NO | 8% | 12% | 27% | 27% | 26% | 100% |

*1 = low, 5 = high difficulty

TABLE 26. PROPORTION OF DIFFICULT ANSWERS LARGER THAN PROPORTION OF EASY ANSWERS

|  | % answers at levels 4 or 5 > % answers at 1 or 2? * | | | | | | | | |
|---|---|---|---|---|---|---|---|---|---|
|  | J1 | J2 | J3 | J4 | J5 | J6 | J7 | J8 | ∑* |
| IDK | 1 | 1 | 1 | 1 | 1 | 1 | 1 | 1 | 8 |
| YES | 0 | 0 | 1 | 1 | 1 | 1 | 0 | 0 | 4 |
| NO | 0 | 1 | 1 | 1 | 1 | 1 | 1 | 0 | 6 |

* 1 = true, 0 = false

Our second assumption concerns the correctness of answers from workers who chose IDK answers more frequently. In order to analyze workers in a fair manner, we considered only their first three answers. We categorized workers in three groups based on the proportion of IDK answers. The only significant difference between average correct answers was between workers with zero versus one IDK in their first three answers (p-value<0.05, medium effect size r=-0.48). Hence, the answer to the exploratory question is that, although we have evidence that higher levels of IDK are related to a higher difficulty, we did not find evidence that IDK level is also related to a larger proportion of incorrect answers. Hence, we could not use the IDK level to consistently filter out workers, which implies that the answer for the exploratory question is negative.

#### 5.5.4.5 Could we locate all faults by filtering out workers from certain professions and lower qualification test score?

We filtered workers by profession and by three levels of the score in the qualification test (e.g., above 60%, above 80%, and 100% score). After applying filters, we found three groups of workers who located all faults (Table 27).

TABLE 27 FILTERS BY WORKER SCORE AND PROFESSION

| Filtered out answers* | precision, recall | lines to inspect | workers | answers |
|---|---|---|---|---|
| students scored < 80% | 78%, 68% | 58 | 231 | 386 |
| all students | 81%, 76% | 56 | 316 | 470 |
| all students U non-student scored < 100% | 94%, 65% | 51 | 133 | 226 |

Table 27 shows that most effective subcrowd (higher precision) was also the smallest one. Hence the answer for the exploratory is affirmative.

#### 5.5.4.6 Could we locate all faults by filtering out workers according to qualification test score and the difficulty level of their answers?

We investigated if answers from different levels of difficulty would have a different proportion of correct answers (i.e., accuracy) across worker scores. Hence, we computed the accuracy of answers by worker score and difficulty level (Table 28).

TABLE 28. ANSWER ACCURACY BY DIFFICULTY LEVEL AND WORKER SCORE

| Worker score | Difficulty levels | | | | |
|---|---|---|---|---|---|
|  | 1 | 2 | 3 | 4 | 5 |
| 60% | 68% | 74% | 66% | <u>59%</u> | <u>55%</u> |
| 80% | 85% | 85% | 71% | 61% | <u>53%</u> |
| 100% | 88% | 77% | 63% | 67% | 64% |

The three underlined cells on the top right of Table 28 present the group of workers with lowest average values of answer accuracy. We also confirmed this pattern at the Java method level. These three highlighted cells corresponded to the lowest accuracy answers in seven out of eight Java methods.

Based on these findings, we composed filters that excluded answers with a high level of difficulty for certain worker scores. Table 29 shows the filters that located all eight faults. Note that the smallest subcrowd provided the highest value of precision. We named this subcrowd as "least difficult answers by worker score". Ultimately, the answer for the exploratory question is affirmative.

TABLE 29. FILTER BY ANSWER DIFFICULTY AND WORKER SCORE

| Filtered out answers | precision, recall | lines to inspect | workers | answers |
|---|---|---|---|---|
| ([5] x [all]) | 81%, 57% | 41 | 456 | 1046 |
| ([5] x [60,80]) | 81%, 60% | 41 | 469 | 1176 |
| ([5] x [80]) U ([4,5] x [60]) | 81%, 60% | 41 | 446 | 1124 |
| ([4,5] x [60,80]) | 88%, 64% | 42 | 431 | 1006 |
| ([5] x [100]) U ([4,5] x [60,80]) | 88%, 61% | 42 | 418 | 876 |

#### 5.5.4.7 Could we locate all faults by filtering out workers according to profession and the difficulty of answers?

Following the previous subsection rationale, we computed the accuracy of answers for each combination of profession and difficulty level (Table 30). The three highlighted cells on the top right of Table 30 present the group of workers with the lowest average values of answer accuracy. We also confirmed this pattern at the Java method level. These three highlighted cells corresponded to the lowest accuracy answers in six out of eight Java methods.

TABLE 30. ANSWER ACCURACY FOR DIFFERENT DIFFICULTY LEVEL AND WORKER PROFESSION

| Worker profession | Difficulty levels | | | | |
|---|---|---|---|---|---|
| | 1 | 2 | 3 | 4 | 5 |
| Undergrad. students | 83% | 74% | 67% | 56% | 31% |
| Graduate students | 92% | 87% | 54% | 72% | 53% |
| Hobbyist | 76% | 79% | 67% | 59% | 63% |
| Others | 87% | 73% | 69% | 61% | 67% |
| Professional developers | 82% | 74% | 68% | 60% | 69% |

We evaluated three filters that excluded answers that were considered very difficult by undergraduate and graduate students. Table 31 shows the different filters that located all eight faults. The last row presented the highest precision. Note, however, that for these filters, the highest precision is not related to the smallest subcrowd, but it still related to the fewest lines to inspect. We named this subcrowd as "least difficult answers by worker profession". Ultimately, the answer for the exploratory question is affirmative.

TABLE 31. FILTER BY ANSWER DIFFICULTY AND WORKER PROFESSION

| Filtered out answers | precision, recall | lines to inspect | workers | answers |
|---|---|---|---|---|
| ([3,4,5] ∩ [non-students]) U [students] | 61%,61% | 67 | 184 | 618 |
| ([5] x [grad.]) U ([4,5]x[undergrad.]) | 75%,64% | 44 | 463 | 2941 |
| ([3,4,5] x [students]) | 77%,61% | 45 | 406 | 2045 |
| ([4,5] x [students]) | 81%,61% | 43 | 454 | 2279 |

### 5.5.5 Summary of subcrowds

Many subcrowds outperformed the original crowd both in terms of precision of answers and lines to inspect. Table 32 displays the statistics of all subcrowds sorted by precision and number of workers. Note although five subcrowds (#1,2,3,4,5) could have outperformed the original crowd (#6) in terms of precision, not all of them outperformed the original crowd in the number of lines to inspect. Conversely, looking at lines to inspect, another four subcrowds (#3, 7, 8, 10) outperformed the original crowd by selecting fewer lines to inspect. This suggests a trade-off between the cost to collect answers versus number of lines to inspect. Additionally, the fewest lines to inspect were selected by subcrowds who are similar in size to the original crowd.

TABLE 32. SUMMARY OF BEST SUBCROWD BY EACH FILTER

| # subcrowd | filters applied | precision, recall | lines to inspect | workers | answers |
|---|---|---|---|---|---|
| 1 | non-students score = 100% | 94%, 65% | 51 | 133 | 836 |
| 2 | worker score = 100% | 94%, 65% | 51 | 194 | 1121 |
| 3 | least difficult answers by worker score | 88%, 61% | 42 | 418 | 1868 |
| 4 | all non-students | 81%, 76% | 56 | 316 | 1173 |
| 5 | least difficult answers by worker profession | 81%, 61% | 43 | 454 | 2279 |
| 6 | all workers | 79%, 64% | 43 | 497 | 2580 |
| 7 | excluded fastest answers | 78%, 59% | 41 | 488 | 2395 |
| 8 | excluded conditionals > 3 LOC | 77%, 76% | 34 | 494 | 2280 |
| 9 | top 3.3% confidence answers | 71%, 56% | 42 | 412 | 1749 |
| 10 | excluded shortest explanations | 70%, 66% | 57 | 412 | 1879 |

We wonder whether the size of the crowd or the number of answers relate to the outcome in terms lines to inspect. In order to investigate the relation between the number of workers and lines to inspect, we plotted the outcomes of all of the 30 subcrowds (Figure 18). The chart suggests a negative correlation, which we confirmed. The number of workers and lines to inspect are shown to have a strong negative correlation (z = -4.75, p-value <0.001, tau=-0.63). However, the same was true for the number of answers and lines to inspect (z = -4.89, p-value <0.001, tau=-0.64). Since the correlation values are similar, we cannot tell which factor has a larger effect on lines to inspect. Hence, we also computed the correlation between the rate of answers per worker versus lines to inspect, which was also negative (z = -3.28, p-value = 0.001 tau=-0.43) i.e., it seems to be better to ask many workers a few questions than to ask many questions to a few workers.

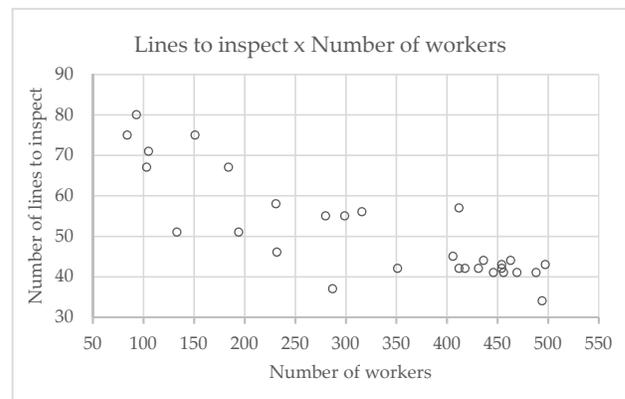

Figure 18 Outcomes from 30 subcrowds in terms of number of workers and lines to inspect

Ultimately, the identified subcrowds are not exhaustive. We found isolated configurations that also located all the eight faults, for instance, selecting only the answers with difficulty level 3 from students or only the answers with difficulty level 2 from professional developers. This shows the importance of looking for patterns that provide context to the data points representing a subcrowd.

## 6 Discussion

The previous section showed that all the eight faults were located by 30 different subcrowds with different degrees of effectiveness, cost, and speed. We now discuss the overall feasibility of the approach to practice, its limitations, and the effects of different crowds. We also look at improvements and possible applications.

### 6.1 Feasibility and its limitations

We discuss feasibility in terms of effectiveness of fault localization, the speed and accuracy of answers, and ease of recruiting enough qualified programmers. While these aspects suggested that the overall approach might be feasible for the simple bug, we identified limitations of cost and scalability.

**Effectiveness.** The subcrowds were effective in a sense that they located all eight faults, which were simple but represented different fault types (Table 1). Moreover, the number of extra lines to inspect was a fraction (11%) of the total lines (211), which is a reasonable proportion of lines for a programmer to inspect to confirm the fault before fixing.

**Speed**. The speed of locating faults showed positive results with regards to the size of the tasks, ability to parallelize work, and worker recruitment. However, the overall speed was limited by the number of questions instantiated and dependency on skillful workers. We discus next each of these feasibility factors.

**Size of tasks.** With regards to the size of tasks, workers completed individual microtasks quickly (5 minutes on average). This reflected the small granularity of tasks, which provided enough information and context for workers to perform tasks quickly, but still effectively (high accuracy). By focusing on answering one question about one code fragment at time, workers had to make fewer assumptions. Since wrong assumptions are known to misguide programmers, particularly novices [103], minimizing the number of assumptions is beneficial to increase speed and accuracy while locating faults.

**Work parallelization.** Another desirable feasibility factor is the ability to parallelize work. Microtasks were performed independently from each other, which in turn enable us to parallelize work among many workers. **Nonetheless, even with high speed of individual answers and work parallelization,** it took the original crowd 77.5 hours to locate all eight faults. One possible explanation for this lag is the time needed to recruit workers. While recruitment was very fast in the first 48 hours (75% of the answers were collected), it took another five days to collect all the remaining answers. One solution is to pre-qualify and train workers for particular tasks, which is already a practice in software testing crowdsourcing platforms [10], [20], [21].

**Worker recruitment. We could recruit enough qualified workers to the number of tasks that we had**. We could qualify enough workers to successfully complete the experiment. The demographics data of these workers contradicts the myth of crowds being composed mostly of amateur workers [104] and the belief that MTurk does not have workers with sufficient software programming skills [105]. Although surprising, this shows that, for similar simple software engineering tasks, a generic platform like MTurk might be suitable to recruit qualified workers.

**Dependency on skilled workers**. However, looking at the subcrowds outcomes (Table 32) we obtained the best results when we considered only non-students with the highest score in their qualification test. Additionally, the high level of participant dropout and qualification failure rates are evidence of the relative scarcity of skillful programmers among the participant workers.

**The design of the microtasks might have mitigated the complexity of the fault localization task.** The source code size and the question type involved in each microtask are the main factors affecting the complexity the microtasks. Hence, we would expect some correlation between these factors and the accuracy of the microtask answers. However, except for the code fragments related to conditional questions, we could not identify any correlation between the accuracy of answers and the size of code fragments or the other question types. Additionally, we could not find any significant correlations between accuracy of answers and the size of the Java methods. This lack of evidence for correlations is can be a positive characteristic of the microtasking approach. Since the source code size and the question type are factors affecting the complexity the microtasks, independence of answer accuracy from these factors suggests that the design of microtasks was effective to mitigate complexity.

**Limited scalability.** Nonetheless, even if we did not find correlations to source code size, there is still a scalability limitation in terms of number of questions asked. We only crowdsourced the failing Java method, i.e., we did not instantiate questions about all the other methods involved in the execution of each failed unit test. Hence, the number of code fragments that would have to be crowdsourced is potentially larger than what we considered in the experiment. We could alleviate that by instantiating only the questions about code fragments that were not executed by the failing unit test.

**Cost limitation**. The scalability is also reflected on the cost to locate each bug. Considering the number of answers and workers of the smallest of the 30 subcrowds (Table 32), each bug cost on average $120 to be located. This is equivalent to a 3-hour work of a professional developer ($38/hour median salary in US [129]). While it might be a reasonable expectation for one programmer to spend 3 hours locating a complex bug, for more simple bugs this might not be cost effective.

One alternative is to pay only Contrary to bug bounty approaches which only pay for the bug found, we paid workers regardless of outcome. This increases overall cost of locating bugs, which for

### 6.2 Effects of different crowds

Besides differences of effectiveness and cost across subcrowds (Table 32), the subcrowds with a larger number of workers produced better results in terms of precision of answers and number of lines to inspect. Moreover, the most qualified groups of workers composed the smaller crowds. This possibly indicates the ideal crowd if we are aiming at high effectiveness at question level and low cost.

However, when speed of results and fewer lines to inspect are more important, subcrowd results (Table 32) showed that larger crowds are probably more indicated. Although larger subcrowds might take longer to recruit, they less strict in terms of worker profile. i.e., more different types of workers can take microtask, hence increasing the task uptake.

### 6.3 Onward

**Rely on fewer answers.** The different subcrowds showed that faults were also located when we considered fewer answers. However, we also showed that a subcrowd with fewer answers produced more false positives. Only when more answers were collected were the number of false positive lines reduced, whereas the number of near positive lines increased (Figure 15). This trade-off could be a decision made by a programmer, who might have preferences in terms of the proportions of false positive lines and near positive lines to inspect.

**Rely on fewer workers.** Selecting workers by certain attribute values might enable requesters to rely on fewer workers to locate faults. However, this also reduces the ability to parallelize work, which might affect the speed of results. Therefore, it would be necessary to compose filters that do not directly impact the number of workers, for instance, filters based on the attributes of answers and question types.

**Allocate more tasks to already qualified workers**. This could speed up task uptake and could be accomplished by adopting mechanisms on top of MTurk (e.g., Turk-prime [110], REACT [111], LegionTools [112]), which would make the task visibility less dependent on MTurk web interface. Additionally, as the same workers perform multiple fault localization microtasks, with time these workers might improve their accuracy. However, this would require some type of feedback mechanism to inform workers about the accuracy of their answers.

**Contemplate more complex bugs**. From the standpoint of the fault localization microtasks utilized in our experiment, we consider two factors affecting bug complexity: the size of the source code to be analyzed by a worker and the amount of supporting information available within the microtask. Take for instance inter-procedural bugs, the source code to be analyzed would comprise program slices that cut across more than one method. In order to focus worker attention on these code fragments, we would need new template questions and a new mechanism to automatically instantiate these questions. Take now bugs that can have different points of a possible null pointer exception. Within each microtask, we could provide the run time state at the time of the failure and allow workers to execute the code fragment (e.g., by extending a cloud IDE).

### 6.4 Applicability

**An alternative approach to bug bounties**. In the case of simple bugs that the current team might not be willing to debug, microtasking could locate faults in a more predictable manner than bug bounties. The reason is that the number of microtasks is known beforehand.

**Flexibility to recruit workers**. Since the fault localization microtasks were completed in minutes, programmers within a company could volunteer to a task force without impacting their project allocation constraints. Additionally, workers from different levels of programming skill could be recruited and can contribute.

**Ad hoc onboarding**. In order to successfully complete a fault localization microtask, each programmer had to acquire minimal understanding about a Java method. While on average the size of workers' explanations about different code fragments within the same Java method did not change significantly (Figure 17 ), workers' speed to provide these explanations increased (Figure 16 ). If this gain in efficiency is a result of workers' improved understanding of the source code, it might be worth investigating microtasking as means of onboarding new programmers to a new software project.

## 7 THREATS TO VALIDITY

The main threats to validity in our study are the conclusion, construct, internal, and external validity threats. Next we discuss their impact and how we mitigated them.

### 7.1 Conclusion Validity

**False positives.** With regards to the statistical methods, we tested for normality (Shapiro-Wilks test) and adopted non-parametric methods when the data failed the normality test. However, normality tests are known to produce false positives and non-parametric tests are also known to produce statistically significant results for large data sets. We mitigated these threats by computing the strength of the effects and providing the rationale for each test. Thereby, others could replicate our results with a different data set.

**False negatives.** The lack of evidence is not an evidence of lack of effect. In order to mitigate this, we disclosed each test that failed to be statistically significant (p-value >= 0.05). Additionally, we also described the situations in which we could not locate all faults, i.e., the exploratory question had a negative answer.

**ad hoc filters.** Defining filters based on the data might lead to problems of over-fitting (i.e., the filter works well only for one data set). We mitigated this by applying filters that were consistent with aspects of the data that are more general, for instance, the distribution of the variables, significant distinctions between means, and correlations between independent and dependent variables.

### 7.2 Construct Validity

**Some of the workers recruited might not be MTurk workers.** The reasons stem from two design constraints. First, we had to utilize MTurk microtasks, called external sites, which enabled us to direct workers to our tool. However, it also did not provide any integration with MTurk system. Second, following IRB constraints, we did not keep an identifiable profile of workers or required login in our tool. Therefore, in order to match MTurk workers' IDs with our internal IDs, we relied on a completion code that we gave to each worker. Workers entered this completion code in the MTurk to prove that they accomplished the task. For workers who did not complete the task, we were not able to make this validation. Nonetheless, we could guarantee that each worker in our experiment is a unique worker because each had a cookie with an identifier saved on their local computer. It is worth mentioning that we had to reject answers from two workers who erased their cookies after failing the qualification test.

**Workers' accuracy might not reflect individual performance.** The reason is that workers might have received help from other workers. We assumed that workers performed relatively simple tasks and expressed their opinions independently. However, it is increasingly recognized that workers collaborate to produce answers [113]. Hence, interpretations about individual worker performance might not be valid. We mitigated that by randomizing the qualification tests and the questions for each new worker. This, however, does not prevent workers from talking to each other or asking help from others. Although that imposes limitations on interpreting individual performance, it does not affect the quality of the outcome of the crowd which might well be an unavoidable reality in crowdsourcing. In a future experiment, we could better evaluate this issue by asking workers which sources of external help they utilized during the task (e.g., web, bug repositories, colleagues, etc.).

**Answer duration might not reflect the time spent on a microtask**. Answer duration shows high variability from seconds to hours for the completion of a single microtask, which makes this measure less trustworthy. Research has also shown that workers tend to manipulate the time of execution, so they are not perceived as being too fast by requesters [35]. One way to cope with this is to add client side scripts to capture worker inactivity as provided in Crowdcode [15].

**Workers' profession might not be consistent with reality.** Since workers' profession was self-declared, workers might have been compelled to exaggerate their programming skills. We mitigate that by stating clearly in the task description that the qualification would be the solely based on a programming test. Nonetheless, based on the differences in qualification test scores among professions (section 4.2), we believe that worker profession is trustworthy. For instance, professional developers were more frequently approved in the qualification tests and quit less often than any other profession.

### 7.3 Internal Validity

**The proportion of correct answers might not be a direct consequence of the filters applied to the data.** Other factors might be influencing the patterns we identified. Since we did not run a controlled experiment, we could not ascertain the exact conditions each worker had in performing the task. For instance, workers were not constrained from looking for help from other workers or from searching the web for hints about the code they were tackling. However, we designed and implemented different mitigation measures. The worker could take the test only once. The worker had a limited time to complete each HIT (2 hours). Additionally, contrary to source code on Github or Sourceforge, the source code of our selected bugs [24] was not easily accessible via a web search.

**Number of workers recruited at each hour might be random.** In order to mitigate distortions in the profiles of workers and the speed of recruitment, we did not post the study anywhere else but on MTurk. Hence, workers had to be logged into MTurk to discover our HITs. However, MTurk is known by popular and very active forums (e.g., Turker Nation [114], Turkopticon [115]), on which workers quickly share recommendations about available HITs.

**The differences in efficiency of locating faults among distinct Java methods might not be a consequence of the characteristics of these methods (e.g., size, complexity, internal structure).** Instead, differences might be related to whether some Java methods have been delegated to more skillful workers than others. The reason for this is that we did not control HIT acceptance by worker profession or test score. HITs were randomly taken on a first come first serve basis, which is a characteristic of the market nature of MTurk. Nevertheless, except for J1 and J2 methods, Java meth-ods did not show significant differences in terms the average worker score (Wilcoxon test of eight pair-wise tests with Bonferroni correction presented p-values>0.0018). Regarding workers' professions, the Java methods with the lowest precision values (J2, J5, J6, J8 using AM.3 in Figure 11) presented neither the lowest proportion of professional developers or the higher proportion of undergraduate students. Even though this lack of evidence does not guarantee that the effect is not present, it suggests at least that the assignment of HITs did not produce a statistical significant imbalance across Java methods.

**Fault location is not a completely objective information.** One alternative is to consider the actual fault location as the same as the location where the fix was made. However, different programmers might have distinct opinions about where a fix should be made and research has shown that the fix locations do not necessarily correspond to the root-cause of the failure [72], [73]. In order to mitigate the threat of having a non-reproducible gold set of fault locations, we investigate the root cause of each failure by running the respective unit tests. Out of the eight Java methods, only for one (J6) the location of the fix was different from the location of the root-cause because the fault in J6 was related to missing code (i.e., new code was added). For this case, we chose the fault location to be the line of code set a variable to an incorrect state (i.e., the root-cause). This shows that building a golden set of actual fault locations involves understanding the root-cause even when the fix locations are known.

### 7.4 External Validity

**The results might not be fully replicated with a different crowd or with a different set of bugs.** We did not analyze how representative the recruited crowd was of the broad population of crowd workers; however, full disclosure of the demographics data can help to compare crowds among different experiments and crowdsourcing platforms.

**Selected bugs might not be representative of all Java bugs.** We did not analyze in depth how representative the eight bugs are of current OSS bugs. Although the simple bugs selected (Table 1) are related with different source code sizes, types of program statements, types of failure, and types of faults, it is still a small sample of eight bugs. We mitigate this validity threat by utilizing bugs from popular open source projects and providing the source code in the appendix of the paper, so others could replicate our study and compare their results with ours. An additional mitigation is to evaluate the scope of the results for bugs located within a Java method (i.e., single file). A recent study [116] shows that bugs affecting a single file comprise the largest set of fixed bugs – from 20% to 40% of all bugs. Hence, besides being available for reproduction, our bug sections is possibly like a large proportion of existing bugs.

**The programming skills of our crowd might not be representative of the population of software programmers available.** Hence, our crowd might not be able to replicate result for more complex bugs. However, looking at the years of experience of workers in our experiment who self-declared as professional developers or graduate students, their experience is like that reported in recent research that discusses the use of students in software engineering experiments [91].

**Our code coverage was not exhaustive**. Our templated questions did not cover the catch exception statements. However, regarding the four code elements that we contemplated (Table 2), research showed these code elements are usually affected by 26.5% to 58.3% of real bugs [117]. Besides that, our four code elements comprise statements both in the data flow and the control flow, which in turn have been shown to be related to 80% of real bugs [116]. Hence, there was a small risk of ignoring lines of code that might have contained the fault. Nonetheless, next experiments should add new template questions that cover all the lines of code.

## 8 Related Work

Several tools and prototypes exist to generate recommendations for software failures by gleaning data from previously executed tasks, for instance, Help-MeOut [118], Crowd::Debug [119], BlueFix [120], and BugFix [122] rely on logs from past programmers to generate fix recommendations for a current failure faced by a new programmer. Other approaches rely on StackOverflow to suggest solutions to coding errors [122] or insights for bug triage [123]. These tools rely on reusing the outcome of tasks performed by previous workers. Hence, these crowdsourcing tools are complementary to the category of crowdsourcing as an open call for contributions from a larger and undefined group of workers.

Research has also presented a long tradition of adopting questions to locate faults. For instance, algorithmic debugging tools [124], [125] prompt users with questions about whether a program statement was correctly executed. The Whyline [126] allows a programmer to select a variable in a Java program and ask "why" or "why not" questions about a program state. CrowdOracles [127] asks questions about the correctness of a source code assertion in regards to the available documentation (JavaDoc). These tools show how questions were extensively adopted as a means of fault localization.

Concerning human studies, research investigated how programmers collect bug related information [106], [107] and how different information affects the effectiveness of fault localization visible to users [128] and the complexity of the source code [100]. Additionally, while studies showed that programmers ask different types of questions about the source code [108], [109], the questions in Stackoverflow that accompanied by source code tend to be answered quickly and satisfactorily [77].

## 9 Conclusion and Future Work

We view this experiment as a necessary first step. If a crowd could not locate the faults in small code fragments, the problem of fault localization in large code bases would be likely very difficult to achieve with microtasking.

Our focus was to investigate the feasibility of template questions to partition fault localization via microtasks and to obtain good quality input from a crowd. For this reason, we used a limited set of questions designed to cover a limited set of intentionally simple bugs. While each bug was taken from a real system with its development history, each was explicitly confined to a single Java method. Clearly, to tackle more complex bugs, we need to expand the design of the questions and the overall workflow.

We showed that the original crowd and other 30 subcrowds could correctly locate all the eight faults with different levels of effectiveness. In the process, we learned three general lessons. First, how questions distributed as microtasks can successfully partition the fault localization of simple bugs in small Java methods. Second, we showed several relationships among answers, questions, and worker attributes. We described heuristics based on these relationships and applied them to select subcrowds who successfully located the faults. Third, we investigate different mechanism for aggregating workers' answers to predict the location of faults. These outcomes were measured in terms of their efficiency and number of lines numbers of lines to be inspected by a programmer.

Our future work involves studying the feasibility of microtasking for more complex bugs. This would involve dealing with inter-procedural dependencies and side effects (i.e., impact on global variables). Additionally, to scale fault localization over multiple methods at a time, we will investigate more complex workflows, for instance, we could ask questions incrementally, basing them on the answers received, or on the structure of the source code, or even allow the workers to suggest new questions.

**ACKNOWLEDGMENT**

This work was supported by the NSF under grant 1414197 and by the CNPq under grant 202005/2012-7. We would also like to thank Danilo Cardoso, Iago Moreira, and Nathan Martins for helping to code and test the various versions of the crowdsourcing environment.

**Christian M. Adriano** received BS and MS degrees in Computer Engineering from the State University of Campinas and a MS degree in Software Engineering from the University of California Irvine. Christian has maintained Project Management Professional credentials since 2008 and has more than ten years of experience in developing software for the banking sector. His current research interests are empirical software engineering, quantitative research methods, program comprehension, software debugging, and program analysis. He is a member of the IEEE and ACM.

**André van der Hoek** received joint BS and MS degrees in business-oriented computer science from the Erasmus University, Rotterdam, The Netherlands, and the Ph.D. degree in computer science from the University of Colorado at Boulder, Boulder, CO. He is currently the Chair of the Department of Informatics at the University of California, Irvine, CA. He heads the Software Design and Collaboration Laboratory, which focuses on understanding and advancing the roles of design, collaboration, and education in software development. He has served on numerous international program committees, is a member of the editorial board of ACM Transactions on Software Engineering and Methodology, was the program chair of the 2010 ACM SIGSOFT International Symposium on the Foundations of Software Engineering, and the program co-chair of the 2014 International Conference on Software Engineering. He was recognized in 2013 as an ACM Distinguished Scientist, and in 2009, he was a recipient of the Premier Award for Excellence in Engineering Education Courseware.

# APPENDIX – I

TABLE 33. SPEED TO LOCATE FAULTS USING AGGREGATION METHODS AM.1 AND AM2

| cut time (hours) | A* | workers | answers | AM.1 | | AM.2 | |
|---|---|---|---|---|---|---|---|
| | | | | faults located | lines to inspect | faults located | lines to inspect |
| **7.2** | 1 | 35 | 129 | 8 | 44 | 0 | - |
| **10.1** | 2 | 66 | 258 | 6 | 43 | 0 | - |
| **12.8** | 3 | 96 | 387 | 7 | 47 | 0 | - |
| **21.5** | 4 | 128 | 516 | 6 | 45 | 0 | - |
| **27.3** | 5 | 158 | 645 | 8 | 43 | 0 | - |
| **31.4** | 6 | 181 | 774 | 8 | 52 | 0 | - |
| **34.4** | 7 | 210 | 903 | 8 | 49 | 0 | - |
| **40.2** | 8 | 243 | 1032 | 8 | 50 | 0 | - |
| **76.4** | 9 | 267 | 1161 | 8 | 45 | 1 | 13 |
| **77.5** | 10 | 288 | 1290 | 7 | 45 | 3 | 24 |
| **78.1** | 11 | 309 | 1419 | 8 | 43 | 4 | 26 |
| **85** | 12 | 339 | 1548 | 8 | 52 | 6 | 35 |
| **98** | 13 | 363 | 1677 | 8 | 53 | 7 | 40 |
| **108.1** | 14 | 386 | 1806 | 8 | 54 | 7 | 41 |
| **108.6** | 15 | 417 | 1935 | 7 | 42 | 7 | 46 |
| **123.5** | 16 | 446 | 2064 | 7 | 43 | 7 | 49 |
| **131.3** | 17 | 460 | 2193 | 7 | 42 | 7 | 54 |
| **138** | 18 | 480 | 2322 | 7 | 42 | 7 | 54 |
| **149.7** | 19 | 490 | 2451 | 7 | 42 | 7 | 55 |
| **154.7** | 20 | 497 | 2580 | 7 | 41 | 8 | 56 |

*A = answers per question

## APPENDIX – II SOURCE CODE OF FAILING JAVA METHODS UTILIZED IN THE EXPERIMENT

These Java methods were obtained from Defect4J repository. Besides the source code, each Java method is followed by the bug report identifier in the Defect4J repository [24], the unit test assertion and failure message presented to the worker in the microtask.

### J1 Java Method

Figure 19. J1 Method - Joda Time, bug 8

### J2 Java Method

Figure 20. J2 Method – JfreeChart, bug 24

### J3 Java Method

Figure 21. J3 Method – CommonsLang, bug 6

### J5 Java Method

Figure 22. J5 Method - CommonsLang, bug 35

### J6 Java Method

Figure 23. J6 Method – Closure, bug 51

**J4 Java Method**

```
256  private void updateBounds(TimePeriod period, int index) {
257
258      long start = period.getStart().getTime();
259      long end = period.getEnd().getTime();
260      long middle = start + ((end - start) / 2);
261
262      if (this.minStartIndex >= 0) {
263          long minStart = getDataItem(this.minStartIndex).getPeriod()
264              .getStart().getTime();
265          if (start < minStart) {
266              this.minStartIndex = index;
267          }
268      }
269      else {
270          this.minStartIndex = index;
271      }
272
273      if (this.maxStartIndex >= 0) {
274          long maxStart = getDataItem(this.maxStartIndex).getPeriod()
275              .getStart().getTime();
276          if (start > maxStart) {
277              this.maxStartIndex = index;
278          }
279      }
280      else {
281          this.maxStartIndex = index;
282      }
283
284      if (this.minMiddleIndex >= 0) {
285          long s = getDataItem(this.minMiddleIndex).getPeriod().getStart()
286              .getTime();
287          long e = getDataItem(this.minMiddleIndex).getPeriod().getEnd()
288              .getTime();
289          long minMiddle = s + (e - s) / 2;
290          if (middle < minMiddle) {
291              this.minMiddleIndex = index;
292          }
293      }
294      else {
295          this.minMiddleIndex = index;
296      }
297
298      if (this.maxMiddleIndex >= 0) {
299          long s = getDataItem(this.minMiddleIndex).getPeriod().getStart()
300              .getTime();
301          long e = getDataItem(this.minMiddleIndex).getPeriod().getEnd()
302              .getTime();
303          long maxMiddle = s + (e - s) / 2;
304          if (middle > maxMiddle) {
305              this.maxMiddleIndex = index;
306          }
307      }
308      else {
309          this.maxMiddleIndex = index;
310      }
311
312      if (this.minEndIndex >= 0) {
313          long minEnd = getDataItem(this.minEndIndex).getPeriod().getEnd()
314              .getTime();
315          if (end < minEnd) {
316              this.minEndIndex = index;
317          }
318      }
319      else {
320          this.minEndIndex = index;
321      }
322
323      if (this.maxEndIndex >= 0) {
324          long maxEnd = getDataItem(this.maxEndIndex).getPeriod().getEnd()
325              .getTime();
326          if (end > maxEnd) {
327              this.maxEndIndex = index;
328          }
329      }
330      else {
331          this.maxEndIndex = index;
332      }
333
334  }
```

Figure 24. J4 Method – Jfreechart, bug 7

**J7 Java Method**

```
892  /**
893   * <p>Converts an array of <code>Object</code> in to an array of <code>Class</code> objects.
894   * If any of these objects is null, a null element will be inserted into the array.</p>
895   *
896   * <p>This method returns <code>null</code> for a <code>null</code> input array.</p>
897   *
898   * @param array an <code>Object</code> array
899   * @return a <code>Class</code> array, <code>null</code> if null array input
900   * @since 2.4
901   */
902  public static Class<?>[] toClass(Object[] array) {
903      if (array == null) {
904          return null;
905      } else if (array.length == 0) {
906          return ArrayUtils.EMPTY_CLASS_ARRAY;
907      }
908      Class<?>[] classes = new Class[array.length];
909      for (int i = 0; i < array.length; i++) {
910          classes[i] = array[i].getClass();
911      }
912      return classes;
913  }
```

Figure 25. J7 Method – CommonsLang, bug 33

**J8 Java Method**

```
65
66   //-----------------------------------------------------------------
67   /**
68    * <p>Converts a String to a Locale.</p>
69    *
70    * <p>This method takes the string format of a locale and creates the
71    * locale object from it.</p>
72    *
73    * <pre>
74    *   LocaleUtils.toLocale("en")         = new Locale("en", "")
75    *   LocaleUtils.toLocale("en_GB")      = new Locale("en", "GB")
76    *   LocaleUtils.toLocale("en_GB_xxx")  = new Locale("en", "GB", "xxx")   (#)
77    * </pre>
78    *
79    * <p>(#) The behaviour of the JDK variant constructor changed between JDK1.3 and JDK1.4.
80    * In JDK1.3, the constructor upper cases the variant, in JDK1.4, it doesn't.
81    * Thus, the result from getVariant() may vary depending on your JDK.</p>
82    *
83    * <p>This method validates the input strictly.
84    * The language code must be lowercase.
85    * The country code must be uppercase.
86    * The separator must be an underscore.
87    * The length must be correct.
88    * </p>
89    *
90    * @param str  the locale String to convert, null returns null
91    * @return a Locale, null if null input
92    * @throws IllegalArgumentException if the string is an invalid format
93    */
94   public static Locale toLocale(String str) {
95       if (str == null) {
96           return null;
97       }
98       int len = str.length();
99       if (len != 2 && len != 5 && len < 7) {
100          throw new IllegalArgumentException("Invalid locale format: " + str);
101      }
102      char ch0 = str.charAt(0);
103      char ch1 = str.charAt(1);
104      if (ch0 < 'a' || ch0 > 'z' || ch1 < 'a' || ch1 > 'z') {
105          throw new IllegalArgumentException("Invalid locale format: " + str);
106      }
107      if (len == 2) {
108          return new Locale(str, "");
109      } else {
110          if (str.charAt(2) != '_') {
111              throw new IllegalArgumentException("Invalid locale format: " + str);
112          }
113          char ch3 = str.charAt(3);
114          char ch4 = str.charAt(4);
115          if (ch3 < 'A' || ch3 > 'Z' || ch4 < 'A' || ch4 > 'Z') {
116              throw new IllegalArgumentException("Invalid locale format: " + str);
117          }
118          if (len == 5) {
119              return new Locale(str.substring(0, 2), str.substring(3, 5));
120          } else {
121              if (str.charAt(5) != '_') {
122                  throw new IllegalArgumentException("Invalid locale format: " + str);
123              }
124              return new Locale(str.substring(0, 2), str.substring(3, 5), str.substring(6));
125          }
126      }
127  }
```

Figure 26. J8 Method – CommonsLang, bug 54